\documentclass[prl,twocolumn,showpacs,preprintnumbers,amsmath,amssymb,tightenlines,epsfig]{revtex4}
\usepackage{graphicx}
\usepackage{color}
\usepackage[percent]{overpic}
\usepackage{enumerate}
\usepackage{enumerate}
\usepackage[export]{adjustbox}

\newcommand{\ket}[1]{|\,{#1}\,\rangle}

\newcommand{\ssection}[1]{{\noi  \it #1:}}

\newcommand{\expec}[1]{\langle #1 \rangle}

\newcommand{\sub}[2]{{#1}_{\mbox{\!\! \scriptsize #2}}}

\def\noi{\noindent}
\def\beq{\begin{equation}}
\def\eeq{\end{equation}}

\def\CR{\nonumber\\[0.15cm]}

\newcommand{\rref}[1]{Ref.~\cite{#1}}
\newcommand{\fref}[1]{Fig.~\ref{#1}}
\newcommand{\frefp}[2]{Fig.~\ref{#1}~(#2)}

\newcommand{\eref}[1]{Eq.~(\ref{#1})}

\newcommand{\cref}[1]{chapter~\ref{#1}}

\newcommand{\Cref}[1]{Chapter~\ref{#1}}

\newcommand{\bref}[1]{(\ref{#1})}

\usepackage{ulem}  
\begin{document}

\title{Hyper-entangling mesoscopic bound states}
\author{Aparna Sreedharan}
\affiliation{Department of Physics, Indian Institute of Science Education and Research (IISER), Bhopal, Madhya Pradesh 462066, India}
\author{Sridevi Kuriyattil}
\affiliation{Department of Physics, Indian Institute of Science Education and Research (IISER), Bhopal, Madhya Pradesh 462066, India}
\author{Sebastian~W\"uster}
\affiliation{Department of Physics, Indian Institute of Science Education and Research (IISER), Bhopal, Madhya Pradesh 462066, India}
\email{sebastian@iiserb.ac.in}
\begin{abstract}
We predict hyper-entanglement generation during binary scattering of mesoscopic bound states, solitary waves in Bose-Einstein condensates containing thousands of identical Bosons. 
The underlying many-body Hamiltonian must not be integrable, and the pre-collision quantum state of the solitons fragmented.
Under these conditions, we show with pure state quantum field simulations that the post-collision state will be hyper-entangled in spatial degrees of freedom and atom number within solitons, for realistic parameters. The effect links aspects of non-linear systems and quantum-coherence and the entangled post-collision state challenges present entanglement criteria for identical particles. Our results are based on simulations of colliding quantum solitons in a quintic interaction model beyond the mean-field, using the truncated Wigner approximation.
\end{abstract}
\maketitle
%
\ssection{Introduction}
%
Quantum mechanics is fundamentally irreconcilable with classical notions such as local realism due to entanglement \cite{EPR_Main_PhysRev,aharonov:bohm:EPR,bell:theorem,Aspect_EPR_experiment}. Seminal explorations were based on pairs of particles originating from a common source, such as a decaying compound particle \cite{aharonov:bohm:EPR} or nonlinear optical processes \cite{Aspect_EPR_experiment}. Similarly, the common source can be a scattering or collision event \cite{Law_entanglecoll_PhysRevA,Jaksch_entangleatomcoll_PhysRevLett}, entangling two earlier separable entities. A single collision then allows a controlled inspection of how interactions entangle complex objects with their surroundings and thus lead to decoherence \cite{schloss_decoherence,Schlosshauer_decoherence_review}.
For projectiles with multiple degrees of freedoms (DGFs), entanglement will in general involve all of these.

Simultaneous entanglement in multiple DGFs has been termed hyper-entanglement \cite{kwiat_hyperentangled_JModOpt}, and can outperform single DGF entanglement for certain tasks in quantum communication \cite{Sheng_PRL_commun1,Sheng_PRA_commun2} and computation \cite{Kwiat_embedded_bell_state_PRA,schuck_prl_bell} as well as quantum cryptography and teleportation \cite{Walborn_hyper_bell_state_PRA,Sheng_PRA_teleport}. It also is of fundamental interest for explorations of the quantum-classical transition such as the generation of exotic mesoscopically entangled states \cite{Li_prl_hyper,Gao_hypercat_NatPhys}, which we explore here.
 
\begin{figure}[htb]
\includegraphics[width=1.0\columnwidth]{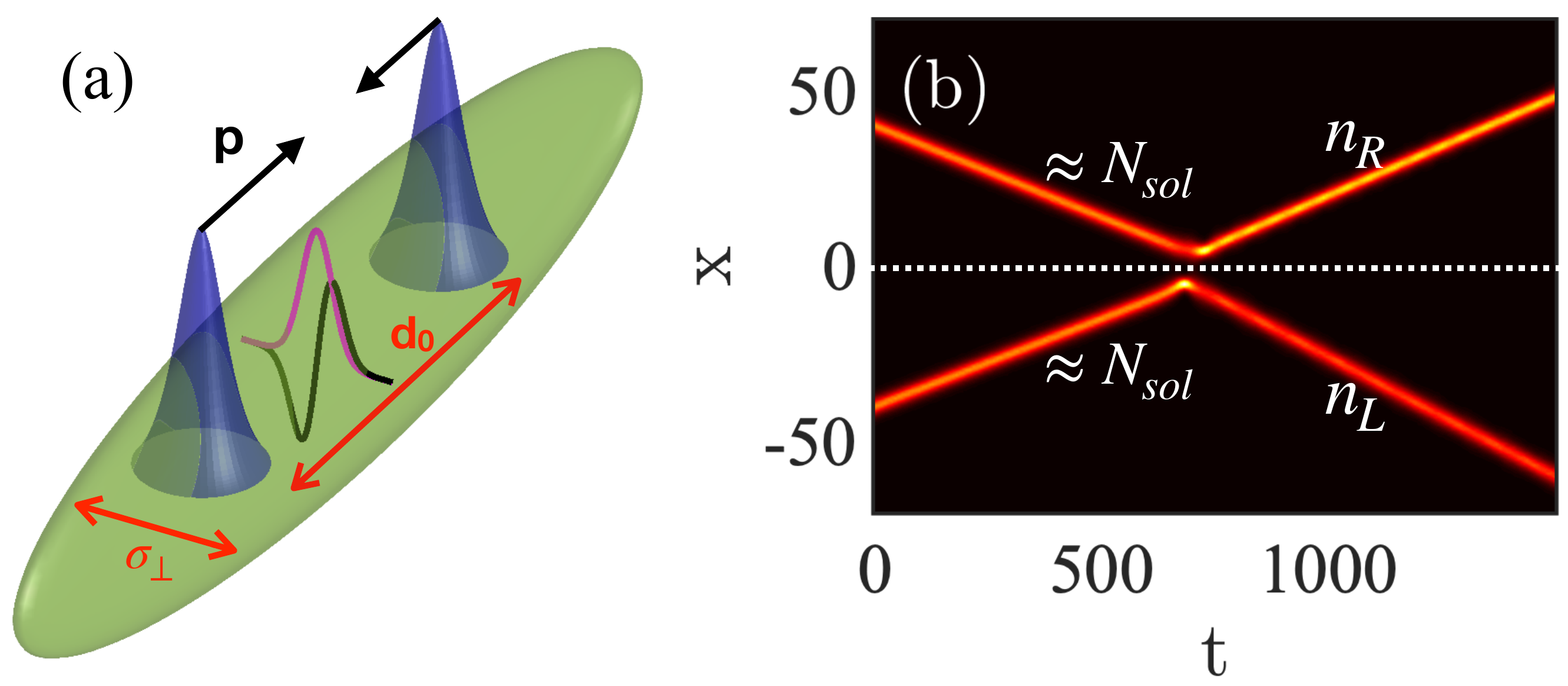}
\caption{\label{sketch} (a) Sketch of soliton collisions. Two solitary waves (blue) in an elongated cigar-shaped trap (green) with transverse oscillator length $\sigma_{\perp}$, initially separated by a distance $d_0$, will collide due to an initial momentum $p$. For a 1D description, the radial wavefunction is fixed in the transverse ground state (violet line), 
but virtual transitions to excited transverse modes (black line) are taken into account through quintic interactions. (b) Total stochastic density $|\phi_{W}(x,t)|^{2}$ of colliding solitons in a single exemplary TWA trajectory (black, zero; bright, high), with $d_0=80$ and $v=0.05$. After the collision, the atom numbers $n_{L}$ in the left and $n_{R}$ in the right soliton differ from their initial value $\approx\sub{N}{sol}$, hence post-collision velocities also differ. The white-dotted horizontal line marks $x=0$ as a guide to the eye.}
\end{figure}

We show that mesoscopic bound-states of thousands of ultracold Bosons, bright matter wave solitons \cite{li_rev,khay:brighsol,li_exp,gap_exp,jila:solitons,Nguyen_modulinst,Nguyen_solcoll_controlled,Marchant_controlledform,Medley_evapsoliton,Lepoutre_sol_Ka,Everitt_modinst,Mesnarsic_cesiumsol_PhysRevA,McDonald_solitoninterf,Marchant_Quantrefl,Boisse_disordersol_EPL,Pollack_extreme_tunability_PRL,parker2008collisions,parker2009bright}, can hyper-entangle in a single collision. During the collision, atoms coherently transfer between the solitons, if there are effective integrability breaking quintic interactions that arise when taking into account transverse modes in the confining potential as shown in \frefp{sketch}{a} \cite{Muryshev_darksolelong_quintic_PRL,Sinha_solfriction_quintic_PRL,Mazets_breakintegrab_PhysRevLett}. 
The resultant superposition state of different atom numbers within each soliton evolves to also exhibit superpositions of momenta and positions after some free evolution, owing to momentum conservation. All three quantities in one soliton are then entangled with those of the collision partner. Both solitons thus are hyper-entangled in constituent number and momentum.

 As opposed to many other carriers of hyper-entanglement, e.g.~\cite{Maximilian_prl_dot, wang_prl_qubit, schuck_prl_bell, ciampini2016path}, the size of a soliton can be continuously scaled by varying its  
 constituent atom number $\sub{N}{sol}$, while important tools for the readout of entanglement such as local oscillators remain available 
 \cite{Kheruntsyan_EPR_moldissoc_PhysRevLett,Gross_homodyne_twinbeam}. Our simulations treat pure states only, and the entangled states are not yet nonlocal, 
 but the complex structure of the post-collision quantum state represent an example of hyperentanglement in two continuous variables that challenges existing entanglement criteria.
 Our results are based on the truncated Wigner approximation (TWA) \cite{steel:wigner,Sinatra2001,castin:validity,blair:review}, which has been shown to give good results regarding creation of entanglement and correlations by comparison with exact methods  \cite{olsen:coherent_transport_JPB,deuar_posPcollisions_PRL,Midgley_comparison_moldissoc_PhysRevA,martin2012quantum}. 

Earlier studies of entanglement generation in soliton collisions did not cover hyper-entanglement and atom transfer due to quintic interactions. Instead, aspects explored were fast collisions that preclude atom transfer \cite{Lewenstein_phasekin_entangle}, internal entanglement in soliton breathers \cite{Lai_entanglesol_PhysRevLett,Ng_nonloc_higherorder_PhysRevLett}, slow entanglement buildup through repeat collisions in a trap \cite{Holdaway_entanglesol}, distinguishable solitons \cite{Gertjerenken_cat_coll_PRL} or dark solitons \cite{Mishmash_entangleddarksol_PRL,Katsimiga_darkbright_NJP2017}. In contrast to many of the above, we demonstrate entanglement generation in a single collision under realistic conditions, that match experiments in \rref{Nguyen_solcoll_controlled}.

\ssection{Solitary waves and effective three-body interactions}
%
We consider an ultracold gas of Bosons with mass $m$, which are free to move in the $x$ direction and harmonically confined transverse to that, with 
Hamiltonian $\sub{\hat{H}}{3D}= \int d^3\mathbf{r} \left[ \hat{\Psi}^\dagger
(\mathbf{r} )\left(-\frac{\hbar^2}{2m}\boldsymbol{\nabla}^2+ \frac{1}{2}m \omega_\perp^2 \mathbf{r}_\perp^2\right)\hat{\Psi}(\mathbf{r} )\right.$
$\left.+ \frac{\sub{U}{0}}{2} \hat{\Psi}^\dagger(\mathbf{r} )\hat{\Psi}^\dagger (\mathbf{r} )\hat{\Psi}(\mathbf{r} )\hat{\Psi}(\mathbf{r}) \right]$,
where the field operator $\hat{\Psi}(\mathbf{r})$ annihilates an atom at position $\mathbf{r}=[x,y,z]^T$. Atomic two-body interactions with 
strength $\sub{U}{0}=4 \pi \hbar^{2} a_{s}/m$ are in the three-dimensional (3D) s-wave scattering regime, where the scattering length is tuned negative $a_s<0$ for attractive interactions. The transverse trapping frequency in the plane $\mathbf{r}_\perp=[y,\: z]^T$ is $\omega_{\perp}$.

For extreme transverse confinement, where even microscopic collisions involve only the dimension $x$
because $\hbar \omega_\perp$ by far exceeds all other energy scales, one obtains the integrable Lieb-Liniger-MacGuire (LL) model \cite{McGuire_exactlysolvable_JMP,LL_model_PR}. There, the set of all \emph{individual} atomic momentum magnitudes is conserved \cite{McGuire_exactlysolvable_JMP,Holdaway_entanglesol,zhu_manybody_liebliniger_ChinPhysB}. Hence, that model does not capture essential features of the more common quasi-1D setting, on which we focus here, in which transverse dynamics is suppressed for the mean-field, but microscopic atomic collisions do involve all three dimensions. For example, the LL model cannot capture the widening momentum distribution of a repulsive quasi-1D condensate freely expanding in a wave guide, as in \rref{Bongs_waveguide_PhysRevA}. 
  
A more adequate quantum field model of quasi-1D condensates is provided by the Hamiltonian
\begin{align}
\hat{H}&= \int dx \bigg\{ \hat{\Psi}^\dagger
(x)\left[-\frac{\hbar^2}{2m}\frac{\partial^2}{\partial x^2}\right]\hat{\Psi}(x)
\CR
&+ \frac{\tilde{g}_{1D}}{2} \hat{\Psi}^\dagger(x)\hat{\Psi}^\dagger (x)\hat{\Psi}(x)\hat{\Psi}(x)
\CR
&
-\frac{\tilde{g}_2}{3} \hat{\Psi}^\dagger(x)\hat{\Psi}^\dagger (x) \hat{\Psi}^\dagger (x)\hat{\Psi}(x)\hat{\Psi}(x) \hat{\Psi}(x)\bigg\},
\label{Hamiltonian1D}
\end{align}
where $\tilde{g}_{1D}=\sub{U}{0}/(2\pi\sigma_{\perp}^2)$ and $\tilde{g}_{2}=U_\perp/(3\pi^{2}\sigma_{\perp}^4)$ from $U_\perp=72 \ln(4/3)\frac{\hbar^{3} a_{s}^{2} \pi^{2}}{m^{2} \omega_{\perp}}$ are effective one dimensional interaction strengths, using a transverse width $\sigma_{\perp}=\sqrt{\hbar/(m\omega_\perp)}$. 
The self-focussing quintic term $\sim-\tilde{g}_2 < 0$ describes effective three-body collisions that arise when integrating out transverse trap modes \cite{Muryshev_darksolelong_quintic_PRL,Sinha_solfriction_quintic_PRL,Mazets_breakintegrab_PhysRevLett} and enables 
dynamically evolving momentum magnitude distributions by breaking the integrability of the case $\tilde{g}_2=0$.

From \eref{Hamiltonian1D}, one can derive the TWA equations of motion \cite{steel:wigner,blair:review} for the stochastic field $\phi_W(x,t)$ as:
\begin{equation}
\begin{split}
 \label{TWA_equation}
i \frac{\partial}{\partial t}\phi_W=  \bigg[-\frac{1}{2} \frac{\partial^2}{\partial x^2} + g_{1D}{(|\phi_W|}^2 - \delta_c)&\\ -q_{2}\big(|\phi_W|^4-2|\phi_W|^2  \delta_c+ \delta_c^{2}\big) \bigg]\phi_W,
\end{split}
\end{equation}   
We now use dimensionless variables, by rescaling $\phi_W \rightarrow \phi_W \sqrt{D}$, $x\rightarrow x/D$,  $t\rightarrow t/T$, where $D$=$\sigma_\perp$ and $T=\omega_{\perp}^{-1}$. The dimensionless interaction constants then take the form: $g_{1D}=2 a_{s}/\sigma_{\perp}$ and $q_{2}=24 \ln[\frac{4}{3} ]a_{s}^{2}/\sigma_\perp^2$.
In \eref{TWA_equation}, $\delta_c=\delta_c(x,x)$ is based on a restricted basis commutator \cite{norrie:prl,norrie:long}, which scales as $\delta_c\sim f_{cut} dx^{-1}$, where $dx$ is the grid spacing and $f_{cut}$ the fraction of Fourier space to which we are adding noise. We choose $f_{cut}=1/2$, to be able to check for aliasing. The TWA method becomes stochastic through the initial state 
\begin{equation}
\label{initialstate}
\phi_{W}(x,0) = \phi_0(x) + \frac{1}{\sqrt 2} \zeta(x),
\end{equation}
where $\phi_0(x)$ is the initial mean field wavefunction and $\zeta(x)$ is a complex Gaussian distributed random function with correlations $\overline{\zeta(x)\zeta(x')}$=0 and $\overline{\zeta^*(x)\zeta(x')}=\delta_c(x,x')$ 
$=\sum_\ell u_\ell(x)u_\ell^*(x')$ \cite{norrie:thesis}. The index $\ell$ numbers a plane wave basis $u_\ell=e^{i k_\ell x}/{\sqrt{\cal V}}$ with normalisation volume ${\cal V}$. The overline denotes the stochastic average, 
which is used to sample quantum correlations, such as $\expec{\hat{\Psi}^\dagger(x)\hat{\Psi}(x')}=\overline{\phi_{W}^*(x)\phi_{W}(x')}-\delta_c(x,x')/2$ \cite{book:qn}.

\ssection{Solitons with quintic nonlinearity}
%
Keeping the quintic term in \bref{TWA_equation} but skipping commutators and initial quantum noise $\zeta$  in \bref{initialstate}, we reach the quintic 
Gross-Pitaevskii-equation (GPE) describing the mean-field. Its solitons and their collisions are discussed in \cite{Khaykovich_Malomed_quinticsol_PRA,Jisha:15_solitons_josab,kivshar2003optical,malomed_stable,nath_quintic,lee2004quantum,baizakov2019effect}. The soliton mean-field wavefunction is
\begin{align}
\label{quintic_solmodes}
\phi(x) = \left(\frac{3}{4q_2}\right)^{1/4}\sqrt\frac{- 4\mu}{\sqrt {g^{2}-4 \mu }\cosh (2\sqrt{-2 \mu} x)+g},
\end{align}
using $g=-g_{1D}\sqrt{3/q_{2}}/2$. The chemical potential $\mu<0$ fixes the atom number per soliton $\sub{N}{sol}$ \cite{Khaykovich_Malomed_quinticsol_PRA}, and in the limit $q_2\rightarrow 0 $, \eref{quintic_solmodes} reduces to the usual sech shape.

To study collisions in the mean-field, one starts with a soliton pair on collision course, separated by $d_0$, 
\begin{align}
\phi_0(x) &=L(x)e^{ikx} +e^{i\varphi} R(x)e^{-ikx},
\label{soliton_pair_meanfield}
\end{align}
with left and right soliton modes $L(x)=\phi(x-d_0/2)$, $R(x)=\phi(x+d_0/2)$, $k$ the initial wave number of the moving soliton and $\varphi$ the initial relative phase. Collisions usually appear attractive for $\varphi = 0$ and repulsive for $\varphi = \pi$, as for solitons in the basic cubic model \cite{gordon_forces,stoof_solitons}. Features that emerge exclusively for $q_2\neq 0 $ are symmetry breaking in collisions for $0<\varphi<\pi$ and
mergers of two solitons for slow collisions \cite{Khaykovich_Malomed_quinticsol_PRA}. Symmetry breaking allows the growth of one soliton at the expense of the other, changing its internal energy and thus representing inelastic collisions. Inelastic soliton collisions due to a cubic-quintic nonlinearity have also been extensively studied in non-linear optics \cite{Cowan_QuasiSoliton_CJP_1986,Sergio_Bistable_OC1992,Soneson_quinticOpticalfibres_PhysicaD,Konar_quinticPropagation_CSF2006,Xie2016_BrightSolitons_Optics_OQE,Albuch_SymAsymOptics_MATCOM_2007}.   

We now include quantum correlations beyond the mean-field, using the TWA for parameters close to recent experiments \cite{Nguyen_solcoll_controlled}, with $\sub{N}{sol}=28000$, $g_{1D}=-2.53\times10^{-5}$ and $q_2=1.10\times10^{-9} = \bar{q}_2$, unless otherwise indicated, corresponding to a scattering length $a_s=-0.030$ nm and $\omega_\perp/(2\pi)=254$ Hz, such that our length and timescales are $D=2.38$ $\mu$m and $T=0.62 $ ms. Since a single stochastic trajectory of \bref{TWA_equation} is found from a solution of the GPE with initial noise \bref{initialstate}, quantum field results can be understood from mean-field dynamics discussed in \rref{Khaykovich_Malomed_quinticsol_PRA}, if we consider stochastic initial conditions. The added noise $\zeta(x)$ randomizes the initial relative phases $\varphi$, initial velocities $v=\hbar k /m$ $(\hbar=m=1)$ and individual atom numbers $n_{L,R}$, e.g.~$n_{L}(0)=\int_{-\infty}^0 dx\left[ |\phi_{W}(x,0)|^2 - \delta_c(x,x)/2\right]$. While the noise is weak enough that $n_L\approx n_R\approx \sub{N}{sol}$, the number fluctuations around this value later
cause large phase-fluctuations through phase-diffusion \cite{lewenstein_phasediff} leading to fragmentation \cite{streltsov_frag,Aparna_collisions_beyondMT_PRA}. We focus on collisions of fragmented solitons, such that despite $\varphi=0$ initially, relative phases at the moment of collision are essentially random.

A representative single trajectory for two colliding solitons is shown in \frefp{sketch}{b}, obtained from a numerical solution of \bref{TWA_equation} using the high-level language XMDS \cite{xmds:paper,xmds:docu}. The relative phase here at the collision is $\varphi\approx 0.68 \: \pi $ causing the transfer of $2a\approx6598$ atoms from the left to the right soliton. Relating $\varphi$ and the half number difference $a=(n_R-n_L)/2$ is nontrivial \cite{Khaykovich_Malomed_quinticsol_PRA,Papacharalampous_solcoll_DNLSE}. The heavier soliton subsequently moves slower than the light one, due to momentum conservation, see \frefp{sketch}{b}. A distracting consequence of the initial noise is the randomization of soliton velocities, causing slight variations of the collision time $\sub{t}{coll}$ and collision point. This represents the diffusion of soliton centres of mass (COM) \cite{weiss_CMdiffusion, Cosme_com_motion}, which we remove from the simulations as discussed in the SI.
\begin{figure}[htb]
\includegraphics[width=1\columnwidth]{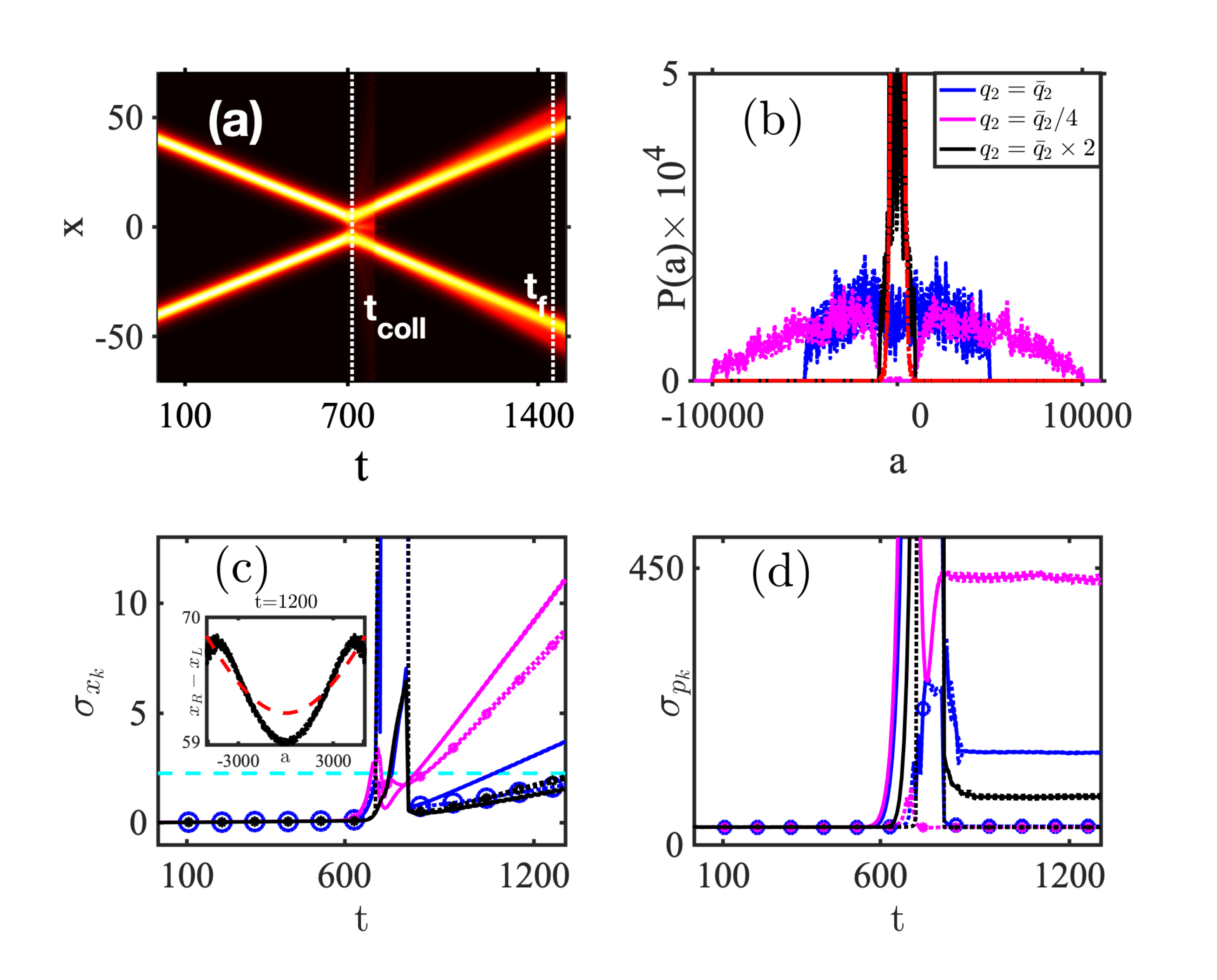}
\caption{\label{collisions} Beyond mean-field collision of fragmented solitons with quintic interactions, generating entanglement.
(a) Square root of mean atomic density $\sqrt{n(x)}$ to emphasize weak features  (black, zero; bright, high). 
Trajectories with a left-right population imbalance exceeding $\sub{A}{cut}$ are dynamically removed. (b) Relative atom number distribution $P(a)$ before collisions (red, dashed) and after collisions at $t_f=1455$ for $q_2=\bar{q}_2$ (blue), $q_2=\bar{q}_2/4$ (magenta), $q_2=\bar{q}_2 \times 2$ (black). Dotted adjacent lines  show the sampling error. (c) Adjusted soliton CM position uncertainty $\Delta[x_R-x_L - \bar{d}(a)]$ (lines with $\circ$) compared to $\Delta_{\Sigma x}$ (solid lines) for the different values of $q_2$ in the same colors as used in (b). The horizontal cyan dashed line shows the width $\xi\approx (2|\mu|)^{-1/2}$ of a single soliton. The inset shows the mean separation $\bar{d}$ at fixed $a$ (black solid) and expectation from \eref{postcollpos} (red dashed) at $t=1200$. (d) Joint momentum uncertainty $\Delta[p_R + p_L]$ (lines with $\circ$) compared to $\Delta_{\Sigma p}$ (solid lines). Dotted lines near solid lines indicate the sampling error. We use $\sub{A}{cut}=(1,5,10) \times 10^3$ for  $q_2/\bar{q}_2=(2,1,1/4)$ respectively.}
 \end{figure}
%

\ssection{Beyond mean-field collisions}
%
For our trajectory averages, we wish to concentrate on binary collisions in the two-mode regime, and remove multi-mode effects such as the excitation of breathers at larger $a$, and soliton mergers at the largest $a$ \cite{Khaykovich_Malomed_quinticsol_PRA}. To this end, trajectories in our stochastic average are dynamically filtered: If the difference of the atom number on the left and right side of the numerical grid exceeds a moderate population imbalance $2a=|n_L - n_R|> \sub{A}{cut}$, indicating a likely merger, trajectories are discarded from that time onwards, such that the resultant ensemble only contains twin soliton final states.

We then find the mean atomic density $n(x)=\expec{\hat{\Psi}^\dagger(x)\hat{\Psi}(x)}=\overline{|\phi_{W}(x)|^2}-\delta_c(x,x)/2$ from an average over initially $\sub{N}{traj}=20000$ individual trajectories similar to the one in \frefp{sketch}{b} and show the result in \frefp{collisions}{a}. We also sample the probability distribution $P(a)$ of the relative atom number $2a$ in \frefp{collisions}{b}, which is initially (red line) Gaussian distributed $P(a)\sim \exp{[-a^2/(2\sigma_a^2)]}$ due to addition of vacuum noise $\zeta(x)$ in \eref{initialstate}, with $\sigma_a\approx\sqrt{2\sub{N}{sol}}$, as expected for two initial coherent state solitons. Since relative phases $\varphi$ at the moment of collision are random and most $\varphi$ cause atom transfer between the solitons \cite{Sreedharan_backforth}, the post collision number difference distribution $P(a)$ (blue and pink lines) can be much wider than the initial one. We explain in \rref{Sreedharan_backforth} why the widening is much enhanced after soliton fragmentation, compared to before, and why the post-collision width does not depend monotonically on the quintic nonlinearity $q_2$. Here we focus on the consequences of the underlying atom transfer, inspecting post-fragmentation collisions only.

Due to momentum conservation, a soliton that has gained atoms at the expense of the other during the collision, must move more slowly afterwards. The resultant link of the atom-number in a soliton and its velocity then gives rise to an increase of the soliton momentum uncertainty in the ensemble, which finally converts into a position uncertainty in the ensemble, as evident by a blurring of the mean atomic density after $\sub{t}{coll}$ in \frefp{collisions}{a}. The atom transfer causing this requires effective three-body interactions: For $q_2=0$ the number distribution $P(a)$ is conserved during collision, as enforced by the integrability of the GPE \cite{Zakharov_solution_NLSE,Sreedharan_backforth}, consequently the density blurring in panel (a) is absent. This reflects that also for the pure 1D quantum field theory \cite{cradle}, there is no atom-transfer between solitons \cite{lai_quantsol_II}, since the rapidity distribution is conserved \cite{McGuire_exactlysolvable_JMP,Holdaway_entanglesol,zhu_manybody_liebliniger_ChinPhysB}. 

\ssection{Hyper-entanglement generation} \label{hyp}
%
We now show that integrability breaking opens the door for hyper-entanglement generation between colliding bright solitons. This is in line with other observations in spin-systems that indicate stronger entanglement generation in non-integrable systems, see e.g.~\cite{Karthik_entangle_chaos_PRA,Deutsch_microsc_entropy_PhysRevE}. Since the model \bref{Hamiltonian1D} is unitary, atom transfer between solitons during the collision is quantum coherent. Schematically, the post collision many-body state $\ket{\sub{\Psi}{pc}}$ can then be written as 
\begin{align}
\ket{\sub{\Psi}{pc}}&=\sum_{a} c_{a}\ket{n(a)_L,v(a)_L}_{L}\otimes  \ket{n(a)_R,v(a)_R}_{R} ,
\label{postcollstate}
\end{align}           
where $\ket{n,v}$ denotes a bound state of $n$ atoms forming a soliton and moving with velocity $v$, and $n(a)_L=\sub{N}{sol}-a$, $n(a)_R=\sub{N}{sol}+a$. Subscripts $L/R$ below the ket distinguish the
left and right soliton. The coefficients $c_{a}\in \mathbb{C}$ are set by the dynamics of the collision and the initial state. Since the TWA  does not provide a many-body quantum state directly, 
we discuss in the following how averages and single trajectories all support that a state of the structure \bref{postcollstate} arises in the simulation.

The initial state is a separable product of the two coherent states for the solitons, and for a separable state the total number $N$ and the relative number $2a$ have the same variance \cite{sup:info}.
Collisions conserve the total number and its variance, while we see in \frefp{collisions}{b} that the variance of the relative number significantly increases. Thus the pure post-collision
state describing atom numbers can no longer be separable \cite{Ng_nonloc_higherorder_PhysRevLett}.

To demonstrate the conversion of number entanglement into momentum entanglement, we use that separable pure states of two solitons must fulfill
\begin{align}
\Delta[p_R+p_L]&=  \sqrt{\Delta[p_R]^2 + \Delta[p_L]^2 }\equiv \Delta_{\Sigma p},
\label{variance_condition_mom}
\end{align}
as shown in the SI. Here, $\Delta[o]$ is the uncertainty (standard deviation) of observable $o$, and $p_L(t)=\int_{-\infty}^0dp [ |\tilde{\phi}_W(p,t)|^2-\tilde{\delta}_c(p,p)/2] p$ and the corresponding quantities for the right soliton, are the centre-of-mass (CM) soliton momentum based on the momentum space wavefunction $\tilde{\phi}_W(p)$ and the restricted basis commutator in momentum space $\tilde{\delta}_c(p,p)$. We show the joint uncertainty $\Delta[p_R+p_L]$ compared with $\Delta_{\Sigma p}$ in \frefp{collisions}{d}, demonstrating that \eref{variance_condition_mom}
is violated for all three cases, hence the momentum state cannot be separable. Since solitons have become entangled in number and momentum, they are hyper-entangled.

Key to the further structure of \eref{postcollstate} is that one can infer both soliton's velocity and then position as a function of relative atom number $a$ from energy and momentum conservation, 
including internal soliton energy but neglecting changes in the mode-shape and the initial number uncertainty. The right soliton moves with dimensionless velocity \cite{sup:info}.
\begin{align}
|v(a)_R|& =  \frac{\sqrt {a-\sub{N}{sol}} \sqrt {a^{2} m( \chi+2\eta \sub{N}{sol}) - p_{0}^{2} \sub{N}{sol}  }}{ \sigma_{\perp} \omega  \: m \sqrt{a \sub{N}{sol}+ \sub{N}{sol}^{2}}},
\label{postcollvelr}
\end{align}
where $\chi$ ($\eta$) parametrise the cubic (quintic) nonlinear energy \cite{footnote:TMMcoeffs}. The left velocity is $|v(a)_L| =  |v(a)_R| (\sub{N}{sol}+a)/(\sub{N}{sol}-a)$. These allow us to predict the expected position of each soliton at time $t$ as
\begin{align}
\bar{x}_{L/R}(a,t)& = \bar{x}_{0L/R}(a)+(t-\sub{t}{coll})v(a)_{L/R},
\label{postcollpos}
\end{align}
and from that their separation $\bar{d}(a)=\bar{x}_R-\bar{x}_L$. One can infer $\sub{t}{coll}= 712.5$ and $\bar{x}_{0L/R}(a)=\pm \sub{d}{min}(a)/2$, with minimal distance $\sub{d}{min}(a)$ from ensemble averages.

We then show in \frefp{collisions}{c} cases where
\begin{align}
\Delta[x_R- x_L - \bar{d}(a)]&<  \sqrt{\Delta [x_R]^2+ \Delta [x_L]^2}\equiv \Delta_{\Sigma x},
\label{variance_condition_pos}
\end{align}
using $x_L(t)=\int_{-\infty}^0dx [ |\phi_W(x,t)|^2-\delta_c(x,x)/2] x$, the stochastic variable representing the CM position of the left soliton within each trajectory, similarly for R.
The sampled distribution of $\bar{d}$ as a function of $a$, and the expected values based on \eref{postcollpos} are shown in the inset. Residual deviations from \eref{postcollpos} (\eref{postcollvelr}) are likely due to the excitation of breathing modes. The figure shows that knowing the position of one soliton and the relative number, we can infer each solitons position better than their overall uncertainty for two cases. For $q_c=2\bar{q}$ this is not possible, since the number distribution has not widened sufficiently. 

Further information that can contribute to the characterisation of the state \bref{postcollstate} are density-density correlations
\begin{equation}
g^{(2)}({x},{x}')=\frac{G^{(2)}({x},{x}')}{n(x)n(x')}=\frac{\langle\hat{\Psi}^\dagger({x}) \hat{\Psi}^\dagger({x}') \hat{\Psi}({x}')\hat{\Psi}({x})\rangle}{\langle\hat{\Psi}^\dagger({x})\hat{\Psi}({x})\rangle \langle\hat{\Psi}^\dagger({x}')\hat{\Psi}({x}')\rangle}.
\label{eq:g2}
\end{equation}
The numerator $G^{(2)}({x},{x}')$ will only be nonzero for two locations $x$, $x'$ where atoms are simultaneously present and is sampled according to the TWA prescription as  
\begin{align}
&G^{(2)}({x},{x}')=\overline{\phi_{W}^*(x')\phi_{W}^*(x)\phi_{W}(x')\phi_{W}(x)}\CR
&-\frac{1}{2} \; \overline{\phi_{W}^*(x)\phi_{W}(x)} \; \delta_{c}(x',x')-\frac{1}{2} \; \overline{\phi_{W}^*(x')\phi_{W}(x')} \; \delta_{c}(x,x)\CR
&-\frac{1}{2} \; \overline{\phi_{W}^*(x')\phi_{W}(x)} \; \delta_{c}(x',x)-\frac{1}{2} \; \overline{\phi_{W}^*(x)\phi_{W}(x')} \; \delta_{c}(x,x')\CR
&+\frac{1}{4} \; \delta_{c}(x,x) \; \delta_{c}(x',x')+\frac{1}{4} \; \delta_{c}(x,x') \; \delta_{c}(x',x).
\label{eq:ddc_corr}
\end{align}
Then, $g^{(2)}({x},{x}')$ is related to the conditional probability to find an atom at $x'$ if one was detected at $x$. 
\begin{figure}[htb]
\includegraphics[width=1\columnwidth]{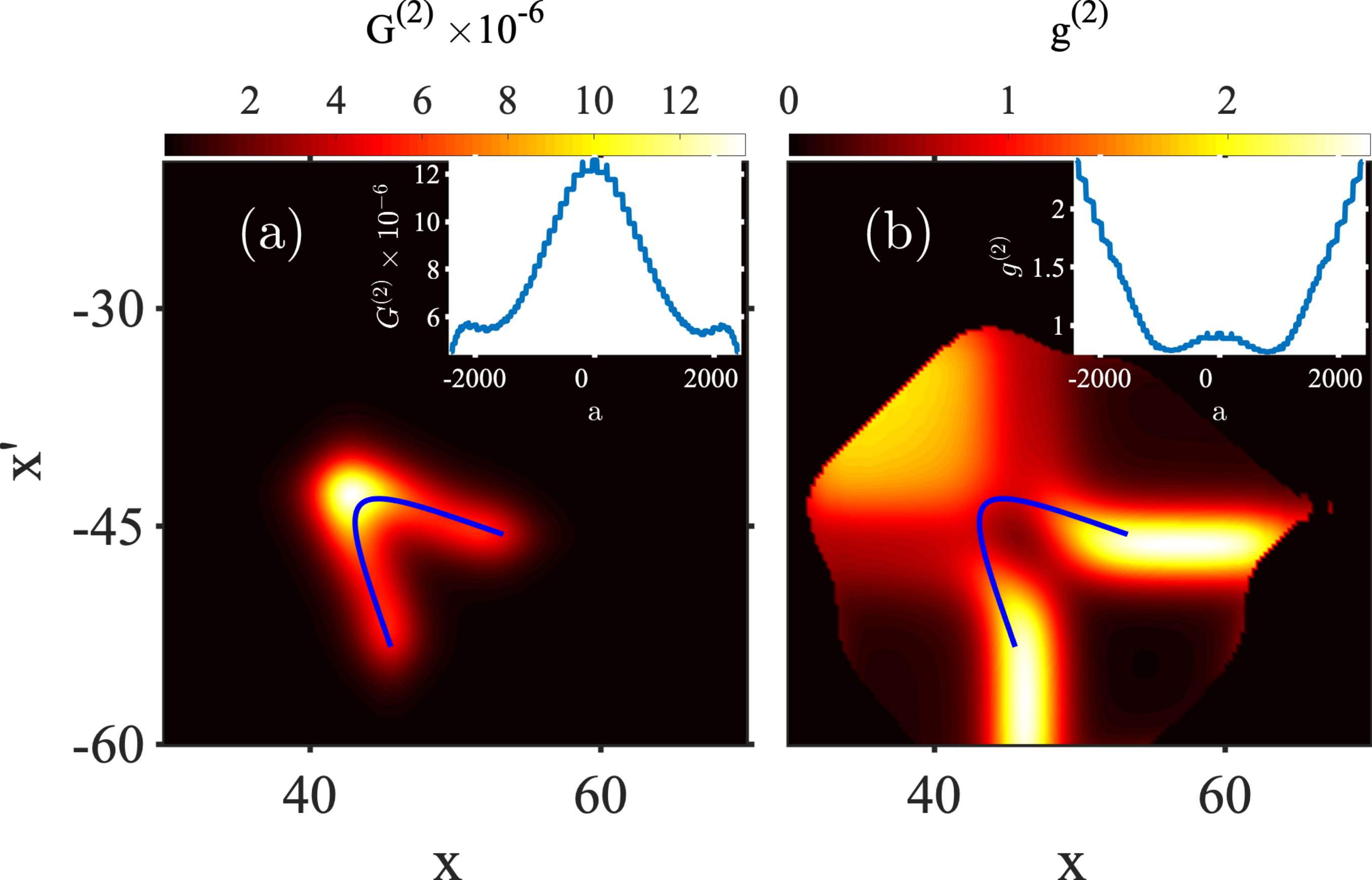}
\caption{\label{correlations} Post collision density-density correlations (a) without normalisation $G^{(2)}$ and (b) with normalisation $g^{(2)}$ at $t^*= t_f$, for the case with $q_2=\bar{q}_2$ shown in \fref{collisions}. We show the blue parametric line $(x , x')$ = $(\bar{x}_R(a,t^*),\bar{x}_L(a,t^*))$ of expected soliton positions in the state \bref{postcollstate} as a function of $a \in [{-2400,2400}]$, an interval matching the width of the distribution in \frefp{collisions}{b}. Insets in both panels show correlations tracing along the blue line.}
\end{figure}
In \fref{correlations} we show the normalized $[g^{(2)}]$ and unnormalised $[G^{(2)}]$ correlations at the time $t_f$ indicated by the white-dotted line in \frefp{collisions}{a}.
Superimposed in blue is the parametric line $(x ,x')$=$(\bar{x}_R(a,t^*),\bar{x}_L(a,t^*))$ indicating at which positions the soliton centres are expected in the state \bref{postcollstate}, for the 
range of transferred atom number $2a$ populated in \frefp{collisions}{b}, with velocities from \eref{postcollvelr}. 
It traces the peak region of $G^{(2)}(x,x')$ well, thus confirming the soliton velocities \bref{postcollvelr} underlying the state \bref{postcollstate}. For most of those positions we also find correlations $g^{(2)}>1$, indicating atom bunching. While \eref{postcollstate} and the subsequent discussion are based on a two-body picture, treating each soliton as a composite object with one internal quantum number (the atom number), we actually model a continuous atomic field describing $N=2\sub{N}{sol}$ atoms. From \frefp{correlations}{a}, we can infer that this field describes a quantum state in which those $N$ atoms
are mesoscopically entangled, residing always in either of two solitons, which are themselves delocalised over a space larger than their width, see cyan line in \frefp{collisions}{c}.

\begin{figure}[htb]
\includegraphics[width=1.\columnwidth]{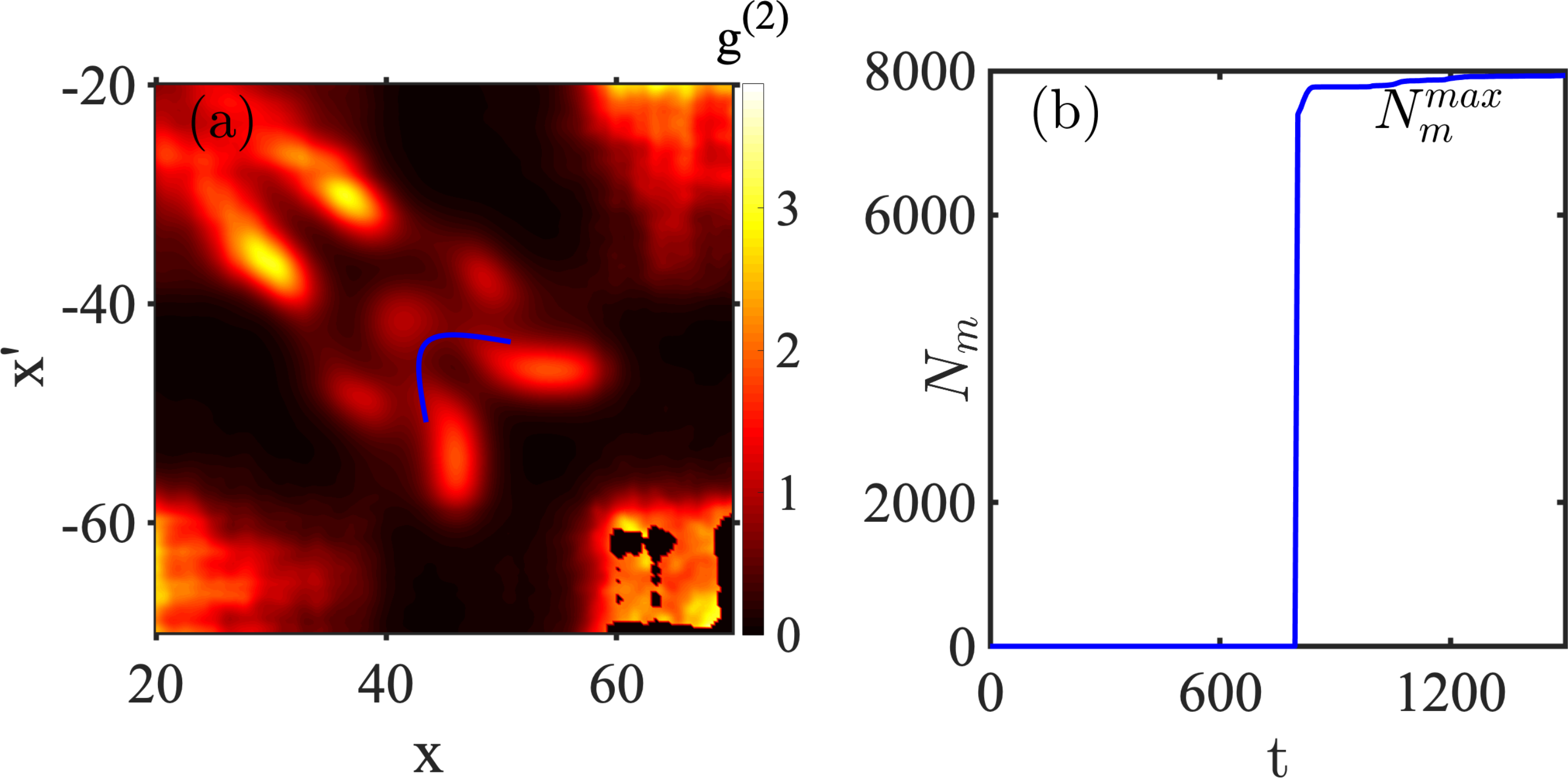}
\caption{\label{fig:merg} (a) Signature of mergers in post collision density-density correlations, for the same scenario as in \frefp{correlations}{b}. (b) The number $N_m$ of discarded trajectories in \fref{correlations} that do not pass the criterion  $|n_L - n_R|>  5000$ as a function of time, with a final number $N_m^{max}\approx{7939}$ of discarded trajectories.}
\end{figure}

Without removing mergers we obtain the correlations shown in \frefp{fig:merg}{a}, which contain the same features as \fref{correlations}, but also additional signatures at coordinates not matching binary soliton collisions described by \eref{postcollpos}. Deviations can occur due to the excitation of breathing modes, which modifies the energy conservation relation. Inspection of single trajectories indicates that these are responsible for the features at $x<40,x'<-40$ in \frefp{correlations}{b} and \frefp{fig:merg}{a}. Even more extreme deviations can be traced back to mergers and radiation from mergers in \fref{fig:merg}. \frefp{fig:merg}{b} shows the number of trajectories $N_{m}$ that are discarded as a function of time when removing mergers, amounting to about $39\%$ of trajectories at the end.

Density-density correlations in Bose-Einstein condensates can now be measured to high precision, and we have shown that these can contain a wealth of information after soliton collisions.
Correlations involving mergers appear at different coordinates $x$, $x'$ than binary collisions, and those regions can provide insight on intra-merger dynamics such as breathing and local entanglement \cite{Ng_nonloc_higherorder_PhysRevLett}.

The TWA method employed here has a good track record in capturing the generation of entanglement \cite{olsen:coherent_transport_JPB,Martin_solinterf_NJP_2012} and correlations \cite{deuar_posPcollisions_PRL,Midgley_comparison_moldissoc_PhysRevA}.
However, it remains approximate, while the complex many-body state \bref{postcollstate} discovered here strongly motivates a future approximation free, many-mode, many-body quantum field method like the Positive-P \cite{drummond1980generalised,deuar2006first} and Gauge-P representations \cite{deuar2006second}, that so far suffer from limited simulation times.

\ssection{Conclusions and Outlook} 
%
 Sampling density-density correlations and joint variances of soliton position and momentum from stochastic quantum field theory, we have provided evidence for the generation of a hyper-entangled state in momentum and atom-number, during condensate soliton collisions in non-integrable scenarios. This relied on the simulation modelling pure states only. Hyper-entanglement generation requires atom transfer between solitons during the collision, enabled by effective three-body collisions that are present naturally in quasi-1D traps \cite{Muryshev_darksolelong_quintic_PRL,Sinha_solfriction_quintic_PRL,Mazets_breakintegrab_PhysRevLett}. Atom transfer may thus serve as an experimental handle to explore these interactions. 
 
The state found in simulations here describes hyperentanglement in two continuous variables, calling for the development of new entanglement criteria 
beyond e.g.~\cite{Zeitler_hyperent_arxiv} and advances in the definition of identical particle entanglement \cite{Benatti_indist_entangle_review}. While the entanglement in soliton positions and momenta shown here does not yet violate Heisenberg limits \cite{Simon_entanglecrit_PhysRevLett, quantele_PRA} and could thus be mimicked classically, the underlying many-body state for many-atoms is clearly mesocopically entangled as in \cite{Cirac_superposBEC_PhysRevA,weiss:solitoncat}, which could ultimately be demonstrated through interference fringes in centre of mass wavefunctions \cite{weiss:solitoncat,moebius:cat}. Fine tuning collisions dynamics could even lead to a hyperentangled version of the kinematic state in the Einstein-Podolsky-Rosen paradox \cite{EPR_Main_PhysRev}, with additional features from many-body physics and number entanglement. 

Finally, the excessively repulsive appearance of soliton collisions in experiments \cite{li_exp,jila:solitons,Nguyen_modulinst,Nguyen_solcoll_controlled} remains unsatisfactorily explained \cite{wuester:collsoll,brand_solitons}. The complex nature of collision dynamics unravelled here might be a key to resolve that.

\acknowledgments
\ssection{Acknowledgements}
We gladly acknowledge the Max-Planck society for funding under the MPG-IISER partner group program, and interesting discussions with Auditya Sharma, Klaus M{\o}lmer, Randall Hulet, Ritesh Pant, Sidharth Rammohan, Shivakant Tiwari and Yash Palan. ASR acknowledges the Department of Science and Technology (DST), New Delhi, India, for the INSPIRE fellowship IF160381. 

\section{Supplemental Information}
\begin{figure}[htb]
\includegraphics[width=0.99\columnwidth]{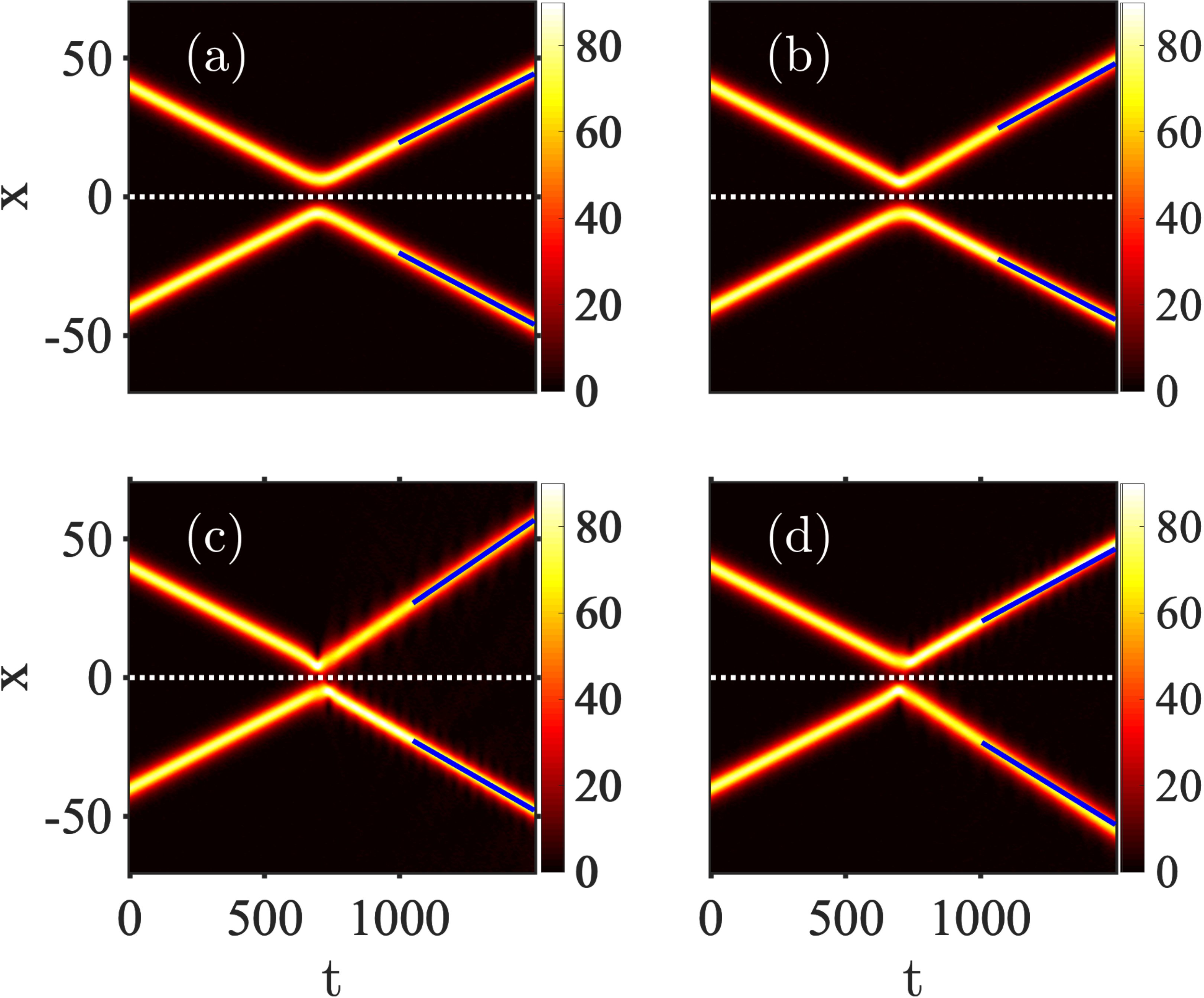}
\caption{\label{singletraj} Additional single trajectories from TWA simulations along with the expected post-collisional velocities. (a-d) show the square root of the total stochastic density $|\phi_{W}(x,t)|^{2}$ (black, zero; bright, high). Superimposed as a blue solid line is the expected trajectory $\bar{x}_L(t-t_{c})$ where $t_{c}$ is the starting point of blue line for the left soliton, with $v(a)_L$ using Eq.~(8) of the main text. The white dotted horizontal line marks x=0 as a guide to the eye.}
\end{figure}

\ssection{Single collision trajectories}
In addition to the single TWA trajectory of Fig.~1(b) in the main article, we show four more trajectories in \fref{singletraj}. The outgoing velocity is compared in detail with the prediction of Eq.~(8) of the main text, for which we extract the transferred atom number $a$ from the individual TWA trajectory.

\ssection{Removing quantum noise on soliton velocities}
%
 The velocity $v_{0}$ intended for the soliton initially, will be slightly changed due to the noise addition in the initial state of TWA trajectories, and becomes $v_{0}+v'$ with small $v'$ causing the post collisional trajectories of soliton centres of mass (COM) diffusive \cite{weiss_CMdiffusion, Cosme_com_motion}.
 However for an analysis of soliton collisions in TWA can be distracting for some parameters, differing from those in the main article, in which case COM diffusion can be removed from simulations as follows.

\begin{figure}[htb]
\includegraphics[width=1.01\columnwidth]{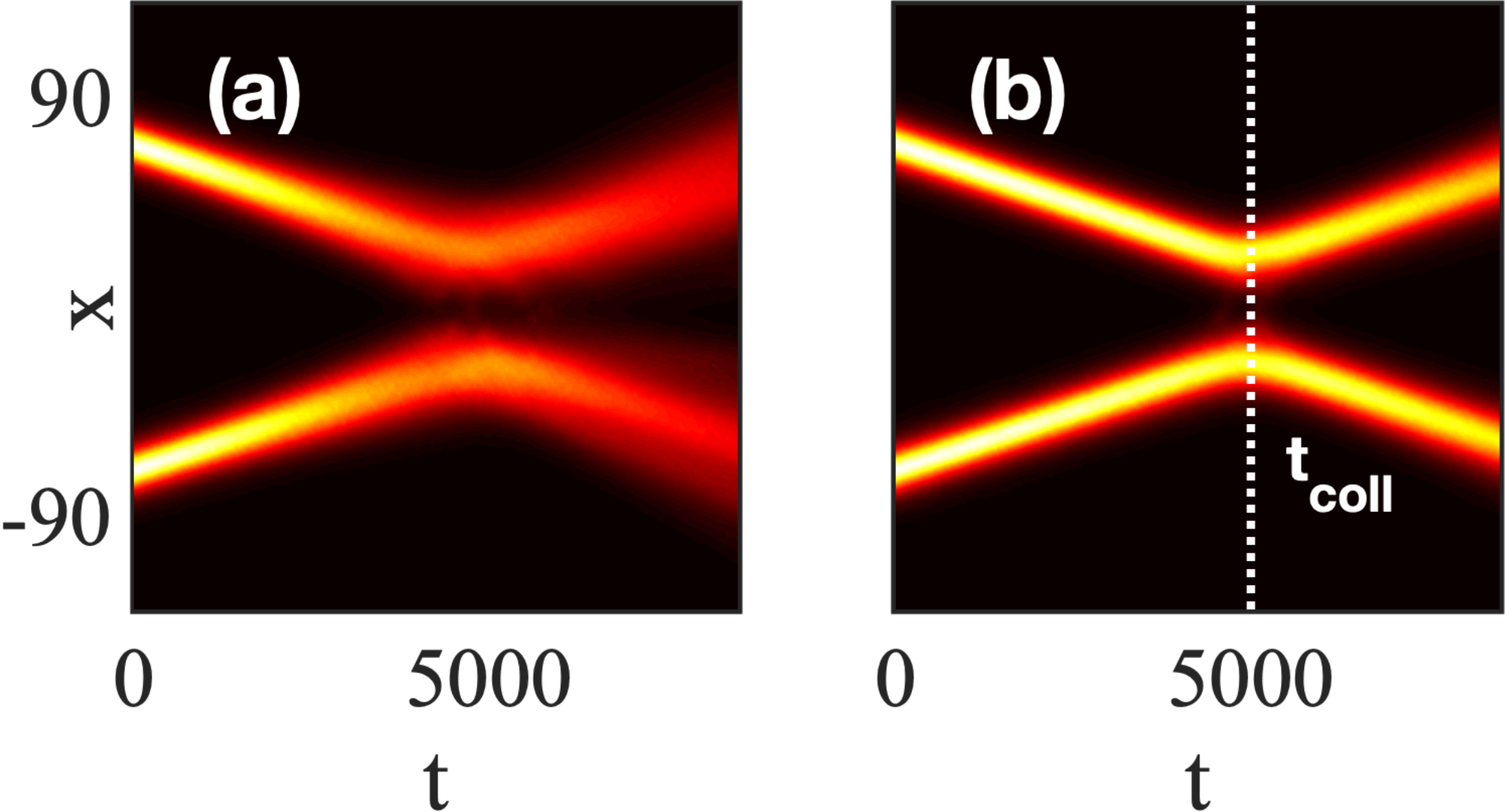}
\caption{\label{fig:veladj} Mean atomic density $n(x)=\expec{\hat{\Psi}^\dagger(x)\hat{\Psi}(x)}$ of colliding solitons of a TWA simulation averaging over $\sub{N}{traj}=100$ (black, zero; bright, high) with $d_0=140$, $\sub{v}{ini}=0.01$ for $\sub{N}{sol}=1000$, $g_{1D}=-2.3\times10^{-4}$, $\sub{q}{2}=9.6 \times 10^{-8}$ corresponding to a scattering length $a_s=-0.15$ nm and $\omega_\perp/(2\pi)=800$ Hz. (a) Without removing quantum noise on the velocity, the collision point is not clearly visible in the average. (b) With adjusting each solitons velocity to the target one, they collide  at $t
_{coll}=5000$ with a clear collision moment shown by the white-dotted line.
}
\end{figure}

To remove quantum noise on soliton velocities, we calculate the local velocity from the stochastic wavefunction $v(x)=\frac{\phi^{*}_W \nabla{\phi_W}}{|\phi_W^{2}|}$ where $|\phi_W^{2}|$ exceeds some density cutoff. We then integrate the function $v(x)$ over the soliton-mode profile $L(x)$ to find $\bar{v}=\int_{-\infty}^{\infty}dx\: \:v(x)\: | L(x)|^{2}$, to re-adjust the soliton velocity by multiplying its stochastic wavefunction by $\exp{[i (v_0 - \bar{v}) x]}$.  Separately applying the procedure to both solitons yields a clear collision point in TWA simulations. \fref{fig:veladj} is an example picked to highlight the utility of velocity adjustment. In contrast, for parameters in the main article which are guided by experiments, the difference is much less severe. 

%
\ssection{Number statistics for a separable two-mode state}
%
Consider the most general pure, separable two-mode state:
\begin{equation}
|\psi \rangle =\sum_{n=0}^{\infty}c_{n} |n \rangle \otimes \sum_{m=0}^{\infty}d_{m} |m \rangle =\sum_{n,m}^{\infty}c_{n}d_{m}|n\: m \rangle.
\label{SI_septwomodestate}
\end{equation}
We define the total- and relative number operators 
\begin{equation}
\hat N=\hat c^{\dagger}\hat c +\hat d^{\dagger} \hat d= \hat N_{L}+\hat N_{R}  
\end{equation}
\begin{equation}
 2\hat a =\hat c^{\dagger}\hat c -\hat d^{\dagger} \hat d =\hat N_{L}-\hat N_{R}  
\end{equation}
The variance of the total number ${\Delta{N}}^{2}$ is then:
\begin{align}
{\Delta{N}}^{2}&=\langle \hat N^{2} \rangle -\langle \hat N \rangle^{2} 
\\&\nonumber=\langle \hat N_{L}^{2} \rangle -\langle \hat N_{L} \rangle^{2}+\langle \hat N_{R}^{2} \rangle -\langle \hat N_{R} \rangle^{2}\\&\nonumber+2(\langle \hat N_{L} \hat N_{R}\rangle-\langle \hat N_{L} \rangle \langle \hat N_{R} \rangle\nonumber) \CR
&={\Delta{N_L}}^{2}+{\Delta{N_R}}^{2},
\end{align}
since $\langle \hat N_{L} \hat N_{R}\rangle=\langle \hat N_{L} \rangle \langle \hat N_{R} \rangle$ for the state \bref{SI_septwomodestate}.

Similarly for the variance of (half of) the relative number ${\Delta{a}}^{2}$, we have
\begin{align}
{\Delta{a}}^{2}&=\langle {\hat a}^{2} \rangle -\langle \hat a \rangle^{2} =\\&\nonumber \frac{1}{4}\bigg[\langle \hat N_{L}^{2} \rangle -\langle \hat N_{L} \rangle^{2}+\langle \hat N_{R}^{2} \rangle -\langle \hat N_{R} \rangle^{2}\\&\nonumber-2(\langle \hat N_{L} \hat N_{R}\rangle-\langle \hat N_{L} \rangle \langle \hat N_{R} \rangle)\bigg] \CR
&=\frac{1}{4}\big[{\Delta{N_L}}^{2}+{\Delta{N_R}}^{2}\big].
\end{align}
We have thus shown that ${\Delta{N}}^{2}=4{\Delta{a}}^{2}$ for a state \bref{SI_septwomodestate},
thus $4{\Delta{a}}^{2}>{\Delta{N}}^{2}$ indicates entanglement for pure states. In our separable initial state  ${\Delta{N}}^{2}=4{\Delta{a}}^{2}$ is fulfilled, 
thus any significant widening of the distribution for $a$ as shown in Fig.~2(b) of the main text indicates entanglement generation for pure states.

\ssection{Position and momentum variances for a separable pure state of two particles}
%
The situation is quite similar for the spatial degrees of freedom for the most general separable pure state of two particles
\begin{equation}
\psi(x_{R},x_{L})=\phi(x_{R})\varphi(x_{L})
\end{equation}
The variance of the momentum sum is then 
\begin{align}
{\Delta(p_{R}+p_{L})}^2&=\langle (p_{R}+p_{L})^{2}\rangle-\langle (p_{R}+p_{L})\rangle^{2}\\&\nonumber=\Delta p_{R}^{2}+\Delta p_{L}^{2}\\&+2(\langle p_{R}p_{L}\rangle-\langle p_{R}\rangle\langle p_{L}\rangle\nonumber) = \Delta p_{R}^{2}+\Delta p_{L}^{2},
\end{align}
while the variance of the position difference, $\Delta(x_{R}-x_{L})^2$
\begin{align}
{\Delta(x_{R}-x_{L})}^2&=\langle (x_{R}-x_{L})^{2}\rangle-\langle (x_{R}-x_{L})\rangle^{2}\\&\nonumber=\Delta x_{R}^{2}+\Delta x_{L}^{2}\\&-2(\langle x_{R}x_{L}\rangle-\langle x_{R}\rangle\langle x_{L}\rangle\nonumber)= \Delta x_{R}^{2}+\Delta x_{L}^{2}.
\end{align}
Thus whenever \emph{either} $${\Delta(p_{R}+p_{L})}< \sqrt{\Delta [p_{R}]^{2}+\Delta [p_{L}]^{2}}\equiv \Delta_{\Sigma p}$$ \emph{or} ${\Delta(x_{R}-x_{L})}< \sqrt{\Delta [x_{R}]^{2}+\Delta [x_{L}]^{2}}\equiv \Delta_{\Sigma x}$, a pure state is entangled.

\bibliography{bibliography}

\providecommand{\noopsort}[1]{}\providecommand{\singleletter}[1]{#1}%
\begin{thebibliography}{110}
\expandafter\ifx\csname natexlab\endcsname\relax\def\natexlab#1{#1}\fi
\expandafter\ifx\csname bibnamefont\endcsname\relax
  \def\bibnamefont#1{#1}\fi
\expandafter\ifx\csname bibfnamefont\endcsname\relax
  \def\bibfnamefont#1{#1}\fi
\expandafter\ifx\csname citenamefont\endcsname\relax
  \def\citenamefont#1{#1}\fi
\expandafter\ifx\csname url\endcsname\relax
  \def\url#1{\texttt{#1}}\fi
\expandafter\ifx\csname urlprefix\endcsname\relax\def\urlprefix{URL }\fi
\providecommand{\bibinfo}[2]{#2}
\providecommand{\eprint}[2][]{\url{#2}}

\bibitem[{\citenamefont{Einstein et~al.}(1935)\citenamefont{Einstein, Podolsky,
  and Rosen}}]{EPR_Main_PhysRev}
\bibinfo{author}{\bibfnamefont{A.}~\bibnamefont{Einstein}},
  \bibinfo{author}{\bibfnamefont{B.}~\bibnamefont{Podolsky}}, \bibnamefont{and}
  \bibinfo{author}{\bibfnamefont{N.}~\bibnamefont{Rosen}},
  \bibinfo{journal}{Phys. Rev.} \textbf{\bibinfo{volume}{47}},
  \bibinfo{pages}{777} (\bibinfo{year}{1935}).

\bibitem[{\citenamefont{Bohm and Aharonov}(1957)}]{aharonov:bohm:EPR}
\bibinfo{author}{\bibfnamefont{D.}~\bibnamefont{Bohm}} \bibnamefont{and}
  \bibinfo{author}{\bibfnamefont{Y.}~\bibnamefont{Aharonov}},
  \bibinfo{journal}{Phys. Rev.} \textbf{\bibinfo{volume}{108}},
  \bibinfo{pages}{1070} (\bibinfo{year}{1957}).

\bibitem[{\citenamefont{Bell}(1964)}]{bell:theorem}
\bibinfo{author}{\bibfnamefont{J.~S.} \bibnamefont{Bell}},
  \bibinfo{journal}{Physics} \textbf{\bibinfo{volume}{1}}, \bibinfo{pages}{195}
  (\bibinfo{year}{1964}).

\bibitem[{\citenamefont{Aspect et~al.}(1982)\citenamefont{Aspect, Grangier, and
  Roger}}]{Aspect_EPR_experiment}
\bibinfo{author}{\bibfnamefont{A.}~\bibnamefont{Aspect}},
  \bibinfo{author}{\bibfnamefont{P.}~\bibnamefont{Grangier}}, \bibnamefont{and}
  \bibinfo{author}{\bibfnamefont{G.}~\bibnamefont{Roger}},
  \bibinfo{journal}{Phys. Rev. Lett.} \textbf{\bibinfo{volume}{49}},
  \bibinfo{pages}{91} (\bibinfo{year}{1982}).

\bibitem[{\citenamefont{Law}(2004)}]{Law_entanglecoll_PhysRevA}
\bibinfo{author}{\bibfnamefont{C.~K.} \bibnamefont{Law}},
  \bibinfo{journal}{Phys. Rev. A} \textbf{\bibinfo{volume}{70}},
  \bibinfo{pages}{062311} (\bibinfo{year}{2004}).

\bibitem[{\citenamefont{Jaksch et~al.}(1999)\citenamefont{Jaksch, Briegel,
  Cirac, Gardiner, and Zoller}}]{Jaksch_entangleatomcoll_PhysRevLett}
\bibinfo{author}{\bibfnamefont{D.}~\bibnamefont{Jaksch}},
  \bibinfo{author}{\bibfnamefont{H.-J.} \bibnamefont{Briegel}},
  \bibinfo{author}{\bibfnamefont{J.~I.} \bibnamefont{Cirac}},
  \bibinfo{author}{\bibfnamefont{C.~W.} \bibnamefont{Gardiner}},
  \bibnamefont{and} \bibinfo{author}{\bibfnamefont{P.}~\bibnamefont{Zoller}},
  \bibinfo{journal}{Phys. Rev. Lett.} \textbf{\bibinfo{volume}{82}},
  \bibinfo{pages}{1975} (\bibinfo{year}{1999}).

\bibitem[{\citenamefont{Schlosshauer}(2007)}]{schloss_decoherence}
\bibinfo{author}{\bibfnamefont{M.~A.} \bibnamefont{Schlosshauer}},
  \emph{\bibinfo{title}{Decoherence: and the quantum-to-classical transition}}
  (\bibinfo{publisher}{Springer Science \& Business Media},
  \bibinfo{year}{2007}).

\bibitem[{\citenamefont{Schlosshauer}(2005)}]{Schlosshauer_decoherence_review}
\bibinfo{author}{\bibfnamefont{M.}~\bibnamefont{Schlosshauer}},
  \bibinfo{journal}{Rev. Mod. Phys.} \textbf{\bibinfo{volume}{76}},
  \bibinfo{pages}{1267} (\bibinfo{year}{2005}).

\bibitem[{\citenamefont{Kwiat}(1997)}]{kwiat_hyperentangled_JModOpt}
\bibinfo{author}{\bibfnamefont{P.~G.} \bibnamefont{Kwiat}},
  \bibinfo{journal}{Journal of Modern Optics} \textbf{\bibinfo{volume}{44}},
  \bibinfo{pages}{2173} (\bibinfo{year}{1997}).

\bibitem[{\citenamefont{Hu et~al.}(2021)\citenamefont{Hu, Huang, Sheng, Zhou,
  Liu, Guo, Zhang, Xing, Huang, Li et~al.}}]{Sheng_PRL_commun1}
\bibinfo{author}{\bibfnamefont{X.-M.} \bibnamefont{Hu}},
  \bibinfo{author}{\bibfnamefont{C.-X.} \bibnamefont{Huang}},
  \bibinfo{author}{\bibfnamefont{Y.-B.} \bibnamefont{Sheng}},
  \bibinfo{author}{\bibfnamefont{L.}~\bibnamefont{Zhou}},
  \bibinfo{author}{\bibfnamefont{B.-H.} \bibnamefont{Liu}},
  \bibinfo{author}{\bibfnamefont{Y.}~\bibnamefont{Guo}},
  \bibinfo{author}{\bibfnamefont{C.}~\bibnamefont{Zhang}},
  \bibinfo{author}{\bibfnamefont{W.-B.} \bibnamefont{Xing}},
  \bibinfo{author}{\bibfnamefont{Y.-F.} \bibnamefont{Huang}},
  \bibinfo{author}{\bibfnamefont{C.-F.} \bibnamefont{Li}},
  \bibnamefont{et~al.}, \bibinfo{journal}{Phys. Rev. Lett.}
  \textbf{\bibinfo{volume}{126}}, \bibinfo{pages}{010503}
  (\bibinfo{year}{2021}).

\bibitem[{\citenamefont{Sheng and Deng}(2010)}]{Sheng_PRA_commun2}
\bibinfo{author}{\bibfnamefont{Y.-B.} \bibnamefont{Sheng}} \bibnamefont{and}
  \bibinfo{author}{\bibfnamefont{F.-G.} \bibnamefont{Deng}},
  \bibinfo{journal}{Phys. Rev. A} \textbf{\bibinfo{volume}{82}},
  \bibinfo{pages}{044305} (\bibinfo{year}{2010}).

\bibitem[{\citenamefont{Kwiat and
  Weinfurter}(1998)}]{Kwiat_embedded_bell_state_PRA}
\bibinfo{author}{\bibfnamefont{P.~G.} \bibnamefont{Kwiat}} \bibnamefont{and}
  \bibinfo{author}{\bibfnamefont{H.}~\bibnamefont{Weinfurter}},
  \bibinfo{journal}{Phys. Rev. A} \textbf{\bibinfo{volume}{58}},
  \bibinfo{pages}{R2623} (\bibinfo{year}{1998}).

\bibitem[{\citenamefont{Schuck et~al.}(2006)\citenamefont{Schuck, Huber,
  Kurtsiefer, and Weinfurter}}]{schuck_prl_bell}
\bibinfo{author}{\bibfnamefont{C.}~\bibnamefont{Schuck}},
  \bibinfo{author}{\bibfnamefont{G.}~\bibnamefont{Huber}},
  \bibinfo{author}{\bibfnamefont{C.}~\bibnamefont{Kurtsiefer}},
  \bibnamefont{and}
  \bibinfo{author}{\bibfnamefont{H.}~\bibnamefont{Weinfurter}},
  \bibinfo{journal}{Phys. Rev. Lett.} \textbf{\bibinfo{volume}{96}},
  \bibinfo{pages}{190501} (\bibinfo{year}{2006}).

\bibitem[{\citenamefont{Walborn et~al.}(2003)\citenamefont{Walborn, P\'adua,
  and Monken}}]{Walborn_hyper_bell_state_PRA}
\bibinfo{author}{\bibfnamefont{S.~P.} \bibnamefont{Walborn}},
  \bibinfo{author}{\bibfnamefont{S.}~\bibnamefont{P\'adua}}, \bibnamefont{and}
  \bibinfo{author}{\bibfnamefont{C.~H.} \bibnamefont{Monken}},
  \bibinfo{journal}{Phys. Rev. A} \textbf{\bibinfo{volume}{68}},
  \bibinfo{pages}{042313} (\bibinfo{year}{2003}).

\bibitem[{\citenamefont{Sheng et~al.}(2010)\citenamefont{Sheng, Deng, and
  Long}}]{Sheng_PRA_teleport}
\bibinfo{author}{\bibfnamefont{Y.-B.} \bibnamefont{Sheng}},
  \bibinfo{author}{\bibfnamefont{F.-G.} \bibnamefont{Deng}}, \bibnamefont{and}
  \bibinfo{author}{\bibfnamefont{G.~L.} \bibnamefont{Long}},
  \bibinfo{journal}{Phys. Rev. A} \textbf{\bibinfo{volume}{82}},
  \bibinfo{pages}{032318} (\bibinfo{year}{2010}).

\bibitem[{\citenamefont{Li et~al.}(2018)\citenamefont{Li, Gessner, Li, and
  Smerzi}}]{Li_prl_hyper}
\bibinfo{author}{\bibfnamefont{Y.}~\bibnamefont{Li}},
  \bibinfo{author}{\bibfnamefont{M.}~\bibnamefont{Gessner}},
  \bibinfo{author}{\bibfnamefont{W.}~\bibnamefont{Li}}, \bibnamefont{and}
  \bibinfo{author}{\bibfnamefont{A.}~\bibnamefont{Smerzi}},
  \bibinfo{journal}{Phys. Rev. Lett.} \textbf{\bibinfo{volume}{120}},
  \bibinfo{pages}{050404} (\bibinfo{year}{2018}).

\bibitem[{\citenamefont{Gao et~al.}(2010)\citenamefont{Gao, Lu, Yao, Xu,
  G{\"u}hne, Goebel, Chen, Peng, Chen, and Pan}}]{Gao_hypercat_NatPhys}
\bibinfo{author}{\bibfnamefont{W.-B.} \bibnamefont{Gao}},
  \bibinfo{author}{\bibfnamefont{C.-Y.} \bibnamefont{Lu}},
  \bibinfo{author}{\bibfnamefont{X.-C.} \bibnamefont{Yao}},
  \bibinfo{author}{\bibfnamefont{P.}~\bibnamefont{Xu}},
  \bibinfo{author}{\bibfnamefont{O.}~\bibnamefont{G{\"u}hne}},
  \bibinfo{author}{\bibfnamefont{A.}~\bibnamefont{Goebel}},
  \bibinfo{author}{\bibfnamefont{Y.-A.} \bibnamefont{Chen}},
  \bibinfo{author}{\bibfnamefont{C.-Z.} \bibnamefont{Peng}},
  \bibinfo{author}{\bibfnamefont{Z.-B.} \bibnamefont{Chen}}, \bibnamefont{and}
  \bibinfo{author}{\bibfnamefont{J.-W.} \bibnamefont{Pan}},
  \bibinfo{journal}{Nature Physics} \textbf{\bibinfo{volume}{6}},
  \bibinfo{pages}{331} (\bibinfo{year}{2010}).

\bibitem[{\citenamefont{Strecker et~al.}(2003)\citenamefont{Strecker,
  Partridge, Truscott, and Hulet}}]{li_rev}
\bibinfo{author}{\bibfnamefont{K.~E.} \bibnamefont{Strecker}},
  \bibinfo{author}{\bibfnamefont{G.~B.} \bibnamefont{Partridge}},
  \bibinfo{author}{\bibfnamefont{A.~G.} \bibnamefont{Truscott}},
  \bibnamefont{and} \bibinfo{author}{\bibfnamefont{R.~G.} \bibnamefont{Hulet}},
  \bibinfo{journal}{New J. Phys.} \textbf{\bibinfo{volume}{5}},
  \bibinfo{pages}{73} (\bibinfo{year}{2003}).

\bibitem[{\citenamefont{Khaykovich et~al.}(2002)\citenamefont{Khaykovich,
  Schreck, Ferrari, Bourdel, Cubizolles, Carr, Castin, and
  Salomon}}]{khay:brighsol}
\bibinfo{author}{\bibfnamefont{L.}~\bibnamefont{Khaykovich}},
  \bibinfo{author}{\bibfnamefont{F.}~\bibnamefont{Schreck}},
  \bibinfo{author}{\bibfnamefont{G.}~\bibnamefont{Ferrari}},
  \bibinfo{author}{\bibfnamefont{T.}~\bibnamefont{Bourdel}},
  \bibinfo{author}{\bibfnamefont{J.}~\bibnamefont{Cubizolles}},
  \bibinfo{author}{\bibfnamefont{L.~D.} \bibnamefont{Carr}},
  \bibinfo{author}{\bibfnamefont{Y.}~\bibnamefont{Castin}}, \bibnamefont{and}
  \bibinfo{author}{\bibfnamefont{C.}~\bibnamefont{Salomon}},
  \bibinfo{journal}{Science} \textbf{\bibinfo{volume}{296}},
  \bibinfo{pages}{1290} (\bibinfo{year}{2002}).

\bibitem[{\citenamefont{Strecker et~al.}(2002)\citenamefont{Strecker,
  Partridge, Truscott, and Hulet}}]{li_exp}
\bibinfo{author}{\bibfnamefont{K.~E.} \bibnamefont{Strecker}},
  \bibinfo{author}{\bibfnamefont{G.~B.} \bibnamefont{Partridge}},
  \bibinfo{author}{\bibfnamefont{A.~G.} \bibnamefont{Truscott}},
  \bibnamefont{and} \bibinfo{author}{\bibfnamefont{R.~G.} \bibnamefont{Hulet}},
  \bibinfo{journal}{Nature} \textbf{\bibinfo{volume}{417}},
  \bibinfo{pages}{150} (\bibinfo{year}{2002}).

\bibitem[{\citenamefont{Eiermann et~al.}(2004)\citenamefont{Eiermann, Anker,
  Albiez, Taglieber, Treutlein, Marzlin, and Oberthaler}}]{gap_exp}
\bibinfo{author}{\bibfnamefont{B.}~\bibnamefont{Eiermann}},
  \bibinfo{author}{\bibfnamefont{T.}~\bibnamefont{Anker}},
  \bibinfo{author}{\bibfnamefont{M.}~\bibnamefont{Albiez}},
  \bibinfo{author}{\bibfnamefont{M.}~\bibnamefont{Taglieber}},
  \bibinfo{author}{\bibfnamefont{P.}~\bibnamefont{Treutlein}},
  \bibinfo{author}{\bibfnamefont{K.-P.} \bibnamefont{Marzlin}},
  \bibnamefont{and} \bibinfo{author}{\bibfnamefont{M.~K.}
  \bibnamefont{Oberthaler}}, \bibinfo{journal}{Phys. Rev. Lett.}
  \textbf{\bibinfo{volume}{92}}, \bibinfo{pages}{230401}
  (\bibinfo{year}{2004}).

\bibitem[{\citenamefont{Cornish et~al.}(2006)\citenamefont{Cornish, Thompson,
  and Wieman}}]{jila:solitons}
\bibinfo{author}{\bibfnamefont{S.~L.} \bibnamefont{Cornish}},
  \bibinfo{author}{\bibfnamefont{S.~T.} \bibnamefont{Thompson}},
  \bibnamefont{and} \bibinfo{author}{\bibfnamefont{C.~E.}
  \bibnamefont{Wieman}}, \bibinfo{journal}{Phys. Rev. Lett.}
  \textbf{\bibinfo{volume}{96}}, \bibinfo{pages}{170401}
  (\bibinfo{year}{2006}).

\bibitem[{\citenamefont{Nguyen et~al.}(2017)\citenamefont{Nguyen, Luo, and
  Hulet}}]{Nguyen_modulinst}
\bibinfo{author}{\bibfnamefont{J.~H.~V.} \bibnamefont{Nguyen}},
  \bibinfo{author}{\bibfnamefont{D.}~\bibnamefont{Luo}}, \bibnamefont{and}
  \bibinfo{author}{\bibfnamefont{R.~G.} \bibnamefont{Hulet}},
  \bibinfo{journal}{Science} \textbf{\bibinfo{volume}{356}},
  \bibinfo{pages}{422} (\bibinfo{year}{2017}).

\bibitem[{\citenamefont{Nguyen et~al.}(2014)\citenamefont{Nguyen, Dyke, Luo,
  Malomed, and Hulet}}]{Nguyen_solcoll_controlled}
\bibinfo{author}{\bibfnamefont{J.~H.~V.} \bibnamefont{Nguyen}},
  \bibinfo{author}{\bibfnamefont{P.}~\bibnamefont{Dyke}},
  \bibinfo{author}{\bibfnamefont{D.}~\bibnamefont{Luo}},
  \bibinfo{author}{\bibfnamefont{B.~A.} \bibnamefont{Malomed}},
  \bibnamefont{and} \bibinfo{author}{\bibfnamefont{R.~G.} \bibnamefont{Hulet}},
  \bibinfo{journal}{Nature Physics} \textbf{\bibinfo{volume}{10}},
  \bibinfo{pages}{918} (\bibinfo{year}{2014}).

\bibitem[{\citenamefont{Marchant et~al.}(2013)\citenamefont{Marchant, Billam,
  Wiles, Yu, Gardiner, and Cornish}}]{Marchant_controlledform}
\bibinfo{author}{\bibfnamefont{A.~L.} \bibnamefont{Marchant}},
  \bibinfo{author}{\bibfnamefont{T.~P.} \bibnamefont{Billam}},
  \bibinfo{author}{\bibfnamefont{T.~P.} \bibnamefont{Wiles}},
  \bibinfo{author}{\bibfnamefont{M.~M.~H.} \bibnamefont{Yu}},
  \bibinfo{author}{\bibfnamefont{S.~A.} \bibnamefont{Gardiner}},
  \bibnamefont{and} \bibinfo{author}{\bibfnamefont{S.~L.}
  \bibnamefont{Cornish}}, \bibinfo{journal}{Nature Comm.}
  \textbf{\bibinfo{volume}{4}}, \bibinfo{pages}{1865} (\bibinfo{year}{2013}).

\bibitem[{\citenamefont{Medley et~al.}(2014)\citenamefont{Medley, Minar, Cizek,
  Berryrieser, and Kasevich}}]{Medley_evapsoliton}
\bibinfo{author}{\bibfnamefont{P.}~\bibnamefont{Medley}},
  \bibinfo{author}{\bibfnamefont{M.~A.} \bibnamefont{Minar}},
  \bibinfo{author}{\bibfnamefont{N.~C.} \bibnamefont{Cizek}},
  \bibinfo{author}{\bibfnamefont{D.}~\bibnamefont{Berryrieser}},
  \bibnamefont{and} \bibinfo{author}{\bibfnamefont{M.~A.}
  \bibnamefont{Kasevich}}, \bibinfo{journal}{Phys. Rev. Lett.}
  \textbf{\bibinfo{volume}{112}}, \bibinfo{pages}{060401}
  (\bibinfo{year}{2014}).

\bibitem[{\citenamefont{Lepoutre et~al.}(2016)\citenamefont{Lepoutre, Fouch\'e,
  Boiss\'e, Berthet, Salomon, Aspect, and Bourdel}}]{Lepoutre_sol_Ka}
\bibinfo{author}{\bibfnamefont{S.}~\bibnamefont{Lepoutre}},
  \bibinfo{author}{\bibfnamefont{L.}~\bibnamefont{Fouch\'e}},
  \bibinfo{author}{\bibfnamefont{A.}~\bibnamefont{Boiss\'e}},
  \bibinfo{author}{\bibfnamefont{G.}~\bibnamefont{Berthet}},
  \bibinfo{author}{\bibfnamefont{G.}~\bibnamefont{Salomon}},
  \bibinfo{author}{\bibfnamefont{A.}~\bibnamefont{Aspect}}, \bibnamefont{and}
  \bibinfo{author}{\bibfnamefont{T.}~\bibnamefont{Bourdel}},
  \bibinfo{journal}{Phys. Rev. A} \textbf{\bibinfo{volume}{94}},
  \bibinfo{pages}{053626} (\bibinfo{year}{2016}).

\bibitem[{\citenamefont{Everitt et~al.}(2017)\citenamefont{Everitt,
  Sooriyabandara, Guasoni, Wigley, Wei, McDonald, Hardman, Manju, Close, Kuhn
  et~al.}}]{Everitt_modinst}
\bibinfo{author}{\bibfnamefont{P.~J.} \bibnamefont{Everitt}},
  \bibinfo{author}{\bibfnamefont{M.~A.} \bibnamefont{Sooriyabandara}},
  \bibinfo{author}{\bibfnamefont{M.}~\bibnamefont{Guasoni}},
  \bibinfo{author}{\bibfnamefont{P.~B.} \bibnamefont{Wigley}},
  \bibinfo{author}{\bibfnamefont{C.~H.} \bibnamefont{Wei}},
  \bibinfo{author}{\bibfnamefont{G.~D.} \bibnamefont{McDonald}},
  \bibinfo{author}{\bibfnamefont{K.~S.} \bibnamefont{Hardman}},
  \bibinfo{author}{\bibfnamefont{P.}~\bibnamefont{Manju}},
  \bibinfo{author}{\bibfnamefont{J.~D.} \bibnamefont{Close}},
  \bibinfo{author}{\bibfnamefont{C.~C.~N.} \bibnamefont{Kuhn}},
  \bibnamefont{et~al.}, \bibinfo{journal}{Phys. Rev. A}
  \textbf{\bibinfo{volume}{96}}, \bibinfo{pages}{041601(R)}
  (\bibinfo{year}{2017}).

\bibitem[{\citenamefont{Me\ifmmode \check{z}\else \v{z}\fi{}nar\ifmmode
  \check{s}\else \v{s}\fi{}i\ifmmode~\check{c}\else \v{c}\fi{}
  et~al.}(2019)\citenamefont{Me\ifmmode \check{z}\else \v{z}\fi{}nar\ifmmode
  \check{s}\else \v{s}\fi{}i\ifmmode~\check{c}\else \v{c}\fi{}, Arh, Brence,
  Pi\ifmmode~\check{s}\else \v{s}\fi{}ljar, Gosar, Gosar,
  \ifmmode~\check{Z}\else \v{Z}\fi{}itko, Zupani\ifmmode~\check{c}\else
  \v{c}\fi{}, and Jegli\ifmmode~\check{c}\else
  \v{c}\fi{}}}]{Mesnarsic_cesiumsol_PhysRevA}
\bibinfo{author}{\bibfnamefont{T.}~\bibnamefont{Me\ifmmode \check{z}\else
  \v{z}\fi{}nar\ifmmode \check{s}\else \v{s}\fi{}i\ifmmode~\check{c}\else
  \v{c}\fi{}}}, \bibinfo{author}{\bibfnamefont{T.}~\bibnamefont{Arh}},
  \bibinfo{author}{\bibfnamefont{J.}~\bibnamefont{Brence}},
  \bibinfo{author}{\bibfnamefont{J.}~\bibnamefont{Pi\ifmmode~\check{s}\else
  \v{s}\fi{}ljar}}, \bibinfo{author}{\bibfnamefont{K.}~\bibnamefont{Gosar}},
  \bibinfo{author}{\bibfnamefont{i.~c.~v.} \bibnamefont{Gosar}},
  \bibinfo{author}{\bibfnamefont{R.}~\bibnamefont{\ifmmode~\check{Z}\else
  \v{Z}\fi{}itko}},
  \bibinfo{author}{\bibfnamefont{E.}~\bibnamefont{Zupani\ifmmode~\check{c}\else
  \v{c}\fi{}}}, \bibnamefont{and}
  \bibinfo{author}{\bibfnamefont{P.}~\bibnamefont{Jegli\ifmmode~\check{c}\else
  \v{c}\fi{}}}, \bibinfo{journal}{Phys. Rev. A} \textbf{\bibinfo{volume}{99}},
  \bibinfo{pages}{033625} (\bibinfo{year}{2019}).

\bibitem[{\citenamefont{McDonald et~al.}(2014)\citenamefont{McDonald, Kuhn,
  Hardman, Bennetts, Everitt, Altin, Debs, Close, and
  Robins}}]{McDonald_solitoninterf}
\bibinfo{author}{\bibfnamefont{G.~D.} \bibnamefont{McDonald}},
  \bibinfo{author}{\bibfnamefont{C.~C.~N.} \bibnamefont{Kuhn}},
  \bibinfo{author}{\bibfnamefont{K.~S.} \bibnamefont{Hardman}},
  \bibinfo{author}{\bibfnamefont{S.}~\bibnamefont{Bennetts}},
  \bibinfo{author}{\bibfnamefont{P.~J.} \bibnamefont{Everitt}},
  \bibinfo{author}{\bibfnamefont{P.~A.} \bibnamefont{Altin}},
  \bibinfo{author}{\bibfnamefont{J.~E.} \bibnamefont{Debs}},
  \bibinfo{author}{\bibfnamefont{J.~D.} \bibnamefont{Close}}, \bibnamefont{and}
  \bibinfo{author}{\bibfnamefont{N.~P.} \bibnamefont{Robins}},
  \bibinfo{journal}{Phys. Rev. Lett.} \textbf{\bibinfo{volume}{113}},
  \bibinfo{pages}{013002} (\bibinfo{year}{2014}).

\bibitem[{\citenamefont{Marchant et~al.}(2016)\citenamefont{Marchant, Billam,
  Yu, Rakonjac, Helm, Polo, Weiss, Gardiner, and Cornish}}]{Marchant_Quantrefl}
\bibinfo{author}{\bibfnamefont{A.~L.} \bibnamefont{Marchant}},
  \bibinfo{author}{\bibfnamefont{T.~P.} \bibnamefont{Billam}},
  \bibinfo{author}{\bibfnamefont{M.~M.~H.} \bibnamefont{Yu}},
  \bibinfo{author}{\bibfnamefont{A.}~\bibnamefont{Rakonjac}},
  \bibinfo{author}{\bibfnamefont{J.~L.} \bibnamefont{Helm}},
  \bibinfo{author}{\bibfnamefont{J.}~\bibnamefont{Polo}},
  \bibinfo{author}{\bibfnamefont{C.}~\bibnamefont{Weiss}},
  \bibinfo{author}{\bibfnamefont{S.~A.} \bibnamefont{Gardiner}},
  \bibnamefont{and} \bibinfo{author}{\bibfnamefont{S.~L.}
  \bibnamefont{Cornish}}, \bibinfo{journal}{Phys. Rev. A}
  \textbf{\bibinfo{volume}{93}}, \bibinfo{pages}{021604(R)}
  (\bibinfo{year}{2016}).

\bibitem[{\citenamefont{Boisse et~al.}({2017})\citenamefont{Boisse, Berthet,
  Fouche, Salomon, Aspect, Lepoutre, and Bourdel}}]{Boisse_disordersol_EPL}
\bibinfo{author}{\bibfnamefont{A.}~\bibnamefont{Boisse}},
  \bibinfo{author}{\bibfnamefont{G.}~\bibnamefont{Berthet}},
  \bibinfo{author}{\bibfnamefont{L.}~\bibnamefont{Fouche}},
  \bibinfo{author}{\bibfnamefont{G.}~\bibnamefont{Salomon}},
  \bibinfo{author}{\bibfnamefont{A.}~\bibnamefont{Aspect}},
  \bibinfo{author}{\bibfnamefont{S.}~\bibnamefont{Lepoutre}}, \bibnamefont{and}
  \bibinfo{author}{\bibfnamefont{T.}~\bibnamefont{Bourdel}},
  \bibinfo{journal}{Eur. Phys. Lett.} \textbf{\bibinfo{volume}{{117}}},
  \bibinfo{pages}{{10007}} (\bibinfo{year}{{2017}}).

\bibitem[{\citenamefont{Pollack et~al.}(2009)\citenamefont{Pollack, Dries,
  Junker, Chen, Corcovilos, and Hulet}}]{Pollack_extreme_tunability_PRL}
\bibinfo{author}{\bibfnamefont{S.~E.} \bibnamefont{Pollack}},
  \bibinfo{author}{\bibfnamefont{D.}~\bibnamefont{Dries}},
  \bibinfo{author}{\bibfnamefont{M.}~\bibnamefont{Junker}},
  \bibinfo{author}{\bibfnamefont{Y.~P.} \bibnamefont{Chen}},
  \bibinfo{author}{\bibfnamefont{T.~A.} \bibnamefont{Corcovilos}},
  \bibnamefont{and} \bibinfo{author}{\bibfnamefont{R.~G.} \bibnamefont{Hulet}},
  \bibinfo{journal}{Phys. Rev. Lett.} \textbf{\bibinfo{volume}{102}},
  \bibinfo{pages}{090402} (\bibinfo{year}{2009}).

\bibitem[{\citenamefont{Parker et~al.}(2008)\citenamefont{Parker, Martin,
  Cornish, and Adams}}]{parker2008collisions}
\bibinfo{author}{\bibfnamefont{N.}~\bibnamefont{Parker}},
  \bibinfo{author}{\bibfnamefont{A.}~\bibnamefont{Martin}},
  \bibinfo{author}{\bibfnamefont{S.}~\bibnamefont{Cornish}}, \bibnamefont{and}
  \bibinfo{author}{\bibfnamefont{C.}~\bibnamefont{Adams}},
  \bibinfo{journal}{Journal of Physics B: Atomic, Molecular and Optical
  Physics} \textbf{\bibinfo{volume}{41}}, \bibinfo{pages}{045303}
  (\bibinfo{year}{2008}).

\bibitem[{\citenamefont{Parker et~al.}(2009)\citenamefont{Parker, Martin,
  Adams, and Cornish}}]{parker2009bright}
\bibinfo{author}{\bibfnamefont{N.}~\bibnamefont{Parker}},
  \bibinfo{author}{\bibfnamefont{A.}~\bibnamefont{Martin}},
  \bibinfo{author}{\bibfnamefont{C.}~\bibnamefont{Adams}}, \bibnamefont{and}
  \bibinfo{author}{\bibfnamefont{S.}~\bibnamefont{Cornish}},
  \bibinfo{journal}{Physica D: Nonlinear Phenomena}
  \textbf{\bibinfo{volume}{238}}, \bibinfo{pages}{1456} (\bibinfo{year}{2009}).

\bibitem[{\citenamefont{Muryshev et~al.}(2002)\citenamefont{Muryshev,
  Shlyapnikov, Ertmer, Sengstock, and
  Lewenstein}}]{Muryshev_darksolelong_quintic_PRL}
\bibinfo{author}{\bibfnamefont{A.}~\bibnamefont{Muryshev}},
  \bibinfo{author}{\bibfnamefont{G.~V.} \bibnamefont{Shlyapnikov}},
  \bibinfo{author}{\bibfnamefont{W.}~\bibnamefont{Ertmer}},
  \bibinfo{author}{\bibfnamefont{K.}~\bibnamefont{Sengstock}},
  \bibnamefont{and}
  \bibinfo{author}{\bibfnamefont{M.}~\bibnamefont{Lewenstein}},
  \bibinfo{journal}{Phys. Rev. Lett.} \textbf{\bibinfo{volume}{89}},
  \bibinfo{pages}{110401} (\bibinfo{year}{2002}).

\bibitem[{\citenamefont{Sinha et~al.}(2006)\citenamefont{Sinha, Cherny,
  Kovrizhin, and Brand}}]{Sinha_solfriction_quintic_PRL}
\bibinfo{author}{\bibfnamefont{S.}~\bibnamefont{Sinha}},
  \bibinfo{author}{\bibfnamefont{A.~Y.} \bibnamefont{Cherny}},
  \bibinfo{author}{\bibfnamefont{D.}~\bibnamefont{Kovrizhin}},
  \bibnamefont{and} \bibinfo{author}{\bibfnamefont{J.}~\bibnamefont{Brand}},
  \bibinfo{journal}{Phys. Rev. Lett.} \textbf{\bibinfo{volume}{96}},
  \bibinfo{pages}{030406} (\bibinfo{year}{2006}).

\bibitem[{\citenamefont{Mazets et~al.}(2008)\citenamefont{Mazets, Schumm, and
  Schmiedmayer}}]{Mazets_breakintegrab_PhysRevLett}
\bibinfo{author}{\bibfnamefont{I.~E.} \bibnamefont{Mazets}},
  \bibinfo{author}{\bibfnamefont{T.}~\bibnamefont{Schumm}}, \bibnamefont{and}
  \bibinfo{author}{\bibfnamefont{J.}~\bibnamefont{Schmiedmayer}},
  \bibinfo{journal}{Phys. Rev. Lett.} \textbf{\bibinfo{volume}{100}},
  \bibinfo{pages}{210403} (\bibinfo{year}{2008}).

\bibitem[{\citenamefont{Prilm\"uller et~al.}(2018)\citenamefont{Prilm\"uller,
  Huber, M\"uller, Michler, Weihs, and Predojevi\ifmmode~\acute{c}\else
  \'{c}\fi{}}}]{Maximilian_prl_dot}
\bibinfo{author}{\bibfnamefont{M.}~\bibnamefont{Prilm\"uller}},
  \bibinfo{author}{\bibfnamefont{T.}~\bibnamefont{Huber}},
  \bibinfo{author}{\bibfnamefont{M.}~\bibnamefont{M\"uller}},
  \bibinfo{author}{\bibfnamefont{P.}~\bibnamefont{Michler}},
  \bibinfo{author}{\bibfnamefont{G.}~\bibnamefont{Weihs}}, \bibnamefont{and}
  \bibinfo{author}{\bibfnamefont{A.}~\bibnamefont{Predojevi\ifmmode~\acute{c}\else
  \'{c}\fi{}}}, \bibinfo{journal}{Phys. Rev. Lett.}
  \textbf{\bibinfo{volume}{121}}, \bibinfo{pages}{110503}
  (\bibinfo{year}{2018}).

\bibitem[{\citenamefont{Wang et~al.}(2018)\citenamefont{Wang, Luo, Huang, Chen,
  Su, Liu, Chen, Li, Fang, Jiang et~al.}}]{wang_prl_qubit}
\bibinfo{author}{\bibfnamefont{X.-L.} \bibnamefont{Wang}},
  \bibinfo{author}{\bibfnamefont{Y.-H.} \bibnamefont{Luo}},
  \bibinfo{author}{\bibfnamefont{H.-L.} \bibnamefont{Huang}},
  \bibinfo{author}{\bibfnamefont{M.-C.} \bibnamefont{Chen}},
  \bibinfo{author}{\bibfnamefont{Z.-E.} \bibnamefont{Su}},
  \bibinfo{author}{\bibfnamefont{C.}~\bibnamefont{Liu}},
  \bibinfo{author}{\bibfnamefont{C.}~\bibnamefont{Chen}},
  \bibinfo{author}{\bibfnamefont{W.}~\bibnamefont{Li}},
  \bibinfo{author}{\bibfnamefont{Y.-Q.} \bibnamefont{Fang}},
  \bibinfo{author}{\bibfnamefont{X.}~\bibnamefont{Jiang}},
  \bibnamefont{et~al.}, \bibinfo{journal}{Phys. Rev. Lett.}
  \textbf{\bibinfo{volume}{120}}, \bibinfo{pages}{260502}
  (\bibinfo{year}{2018}).

\bibitem[{\citenamefont{Ciampini et~al.}(2016)\citenamefont{Ciampini, Orieux,
  Paesani, Sciarrino, Corrielli, Crespi, Ramponi, Osellame, and
  Mataloni}}]{ciampini2016path}
\bibinfo{author}{\bibfnamefont{M.~A.} \bibnamefont{Ciampini}},
  \bibinfo{author}{\bibfnamefont{A.}~\bibnamefont{Orieux}},
  \bibinfo{author}{\bibfnamefont{S.}~\bibnamefont{Paesani}},
  \bibinfo{author}{\bibfnamefont{F.}~\bibnamefont{Sciarrino}},
  \bibinfo{author}{\bibfnamefont{G.}~\bibnamefont{Corrielli}},
  \bibinfo{author}{\bibfnamefont{A.}~\bibnamefont{Crespi}},
  \bibinfo{author}{\bibfnamefont{R.}~\bibnamefont{Ramponi}},
  \bibinfo{author}{\bibfnamefont{R.}~\bibnamefont{Osellame}}, \bibnamefont{and}
  \bibinfo{author}{\bibfnamefont{P.}~\bibnamefont{Mataloni}},
  \bibinfo{journal}{Light: Science \& Applications}
  \textbf{\bibinfo{volume}{5}}, \bibinfo{pages}{e16064} (\bibinfo{year}{2016}).

\bibitem[{\citenamefont{Kheruntsyan et~al.}(2005)\citenamefont{Kheruntsyan,
  Olsen, and Drummond}}]{Kheruntsyan_EPR_moldissoc_PhysRevLett}
\bibinfo{author}{\bibfnamefont{K.~V.} \bibnamefont{Kheruntsyan}},
  \bibinfo{author}{\bibfnamefont{M.~K.} \bibnamefont{Olsen}}, \bibnamefont{and}
  \bibinfo{author}{\bibfnamefont{P.~D.} \bibnamefont{Drummond}},
  \bibinfo{journal}{Phys. Rev. Lett.} \textbf{\bibinfo{volume}{95}},
  \bibinfo{pages}{150405} (\bibinfo{year}{2005}).

\bibitem[{\citenamefont{Gross et~al.}(2011)\citenamefont{Gross, Strobel,
  Nicklas, Zibold, Bar-Gill, Kurizki, and
  Oberthaler}}]{Gross_homodyne_twinbeam}
\bibinfo{author}{\bibfnamefont{C.}~\bibnamefont{Gross}},
  \bibinfo{author}{\bibfnamefont{H.}~\bibnamefont{Strobel}},
  \bibinfo{author}{\bibfnamefont{E.}~\bibnamefont{Nicklas}},
  \bibinfo{author}{\bibfnamefont{T.}~\bibnamefont{Zibold}},
  \bibinfo{author}{\bibfnamefont{N.}~\bibnamefont{Bar-Gill}},
  \bibinfo{author}{\bibfnamefont{G.}~\bibnamefont{Kurizki}}, \bibnamefont{and}
  \bibinfo{author}{\bibfnamefont{M.~K.} \bibnamefont{Oberthaler}},
  \bibinfo{journal}{Nature} \textbf{\bibinfo{volume}{480}},
  \bibinfo{pages}{219} (\bibinfo{year}{2011}).

\bibitem[{\citenamefont{Steel et~al.}(1998)\citenamefont{Steel, Olsen, Plimak,
  Drummond, Tan, Collett, Walls, and Graham}}]{steel:wigner}
\bibinfo{author}{\bibfnamefont{M.~J.} \bibnamefont{Steel}},
  \bibinfo{author}{\bibfnamefont{M.~K.} \bibnamefont{Olsen}},
  \bibinfo{author}{\bibfnamefont{L.~I.} \bibnamefont{Plimak}},
  \bibinfo{author}{\bibfnamefont{P.~D.} \bibnamefont{Drummond}},
  \bibinfo{author}{\bibfnamefont{S.~M.} \bibnamefont{Tan}},
  \bibinfo{author}{\bibfnamefont{M.~J.} \bibnamefont{Collett}},
  \bibinfo{author}{\bibfnamefont{D.~F.} \bibnamefont{Walls}}, \bibnamefont{and}
  \bibinfo{author}{\bibfnamefont{R.}~\bibnamefont{Graham}},
  \bibinfo{journal}{Phys. Rev. A} \textbf{\bibinfo{volume}{58}},
  \bibinfo{pages}{4824} (\bibinfo{year}{1998}).

\bibitem[{\citenamefont{Sinatra et~al.}(2001)\citenamefont{Sinatra, Lobo, and
  Castin}}]{Sinatra2001}
\bibinfo{author}{\bibfnamefont{A.}~\bibnamefont{Sinatra}},
  \bibinfo{author}{\bibfnamefont{C.}~\bibnamefont{Lobo}}, \bibnamefont{and}
  \bibinfo{author}{\bibfnamefont{Y.}~\bibnamefont{Castin}},
  \bibinfo{journal}{Phys. Rev. Lett.} \textbf{\bibinfo{volume}{87}},
  \bibinfo{pages}{210404} (\bibinfo{year}{2001}).

\bibitem[{\citenamefont{Sinatra et~al.}(2002)\citenamefont{Sinatra, Lobo, and
  Castin}}]{castin:validity}
\bibinfo{author}{\bibfnamefont{A.}~\bibnamefont{Sinatra}},
  \bibinfo{author}{\bibfnamefont{C.}~\bibnamefont{Lobo}}, \bibnamefont{and}
  \bibinfo{author}{\bibfnamefont{Y.}~\bibnamefont{Castin}},
  \bibinfo{journal}{J. Phys. B: At. Mol. Opt. Phys.}
  \textbf{\bibinfo{volume}{35}}, \bibinfo{pages}{3599} (\bibinfo{year}{2002}).

\bibitem[{\citenamefont{Blakie et~al.}(2008)\citenamefont{Blakie, Bradley,
  Davis, Ballagh, and Gardiner}}]{blair:review}
\bibinfo{author}{\bibfnamefont{P.}~\bibnamefont{Blakie}},
  \bibinfo{author}{\bibfnamefont{A.}~\bibnamefont{Bradley}},
  \bibinfo{author}{\bibfnamefont{M.}~\bibnamefont{Davis}},
  \bibinfo{author}{\bibfnamefont{R.}~\bibnamefont{Ballagh}}, \bibnamefont{and}
  \bibinfo{author}{\bibfnamefont{C.}~\bibnamefont{Gardiner}},
  \bibinfo{journal}{Advances in Physics} \textbf{\bibinfo{volume}{57}},
  \bibinfo{pages}{363} (\bibinfo{year}{2008}).

\bibitem[{\citenamefont{Olsen}(2014)}]{olsen:coherent_transport_JPB}
\bibinfo{author}{\bibfnamefont{M.~K.} \bibnamefont{Olsen}},
  \bibinfo{journal}{J. Phys. B: At. Mol. Opt. Phys.}
  \textbf{\bibinfo{volume}{47}}, \bibinfo{pages}{095301}
  (\bibinfo{year}{2014}).

\bibitem[{\citenamefont{Deuar and Drummond}(2007)}]{deuar_posPcollisions_PRL}
\bibinfo{author}{\bibfnamefont{P.}~\bibnamefont{Deuar}} \bibnamefont{and}
  \bibinfo{author}{\bibfnamefont{P.~D.} \bibnamefont{Drummond}},
  \bibinfo{journal}{Phys. Rev. Lett.} \textbf{\bibinfo{volume}{98}},
  \bibinfo{pages}{120402} (\bibinfo{year}{2007}).

\bibitem[{\citenamefont{Midgley et~al.}(2009)\citenamefont{Midgley, W\"uster,
  Olsen, Davis, and Kheruntsyan}}]{Midgley_comparison_moldissoc_PhysRevA}
\bibinfo{author}{\bibfnamefont{S.~L.~W.} \bibnamefont{Midgley}},
  \bibinfo{author}{\bibfnamefont{S.}~\bibnamefont{W\"uster}},
  \bibinfo{author}{\bibfnamefont{M.~K.} \bibnamefont{Olsen}},
  \bibinfo{author}{\bibfnamefont{M.~J.} \bibnamefont{Davis}}, \bibnamefont{and}
  \bibinfo{author}{\bibfnamefont{K.~V.} \bibnamefont{Kheruntsyan}},
  \bibinfo{journal}{Phys. Rev. A} \textbf{\bibinfo{volume}{79}},
  \bibinfo{pages}{053632} (\bibinfo{year}{2009}).

\bibitem[{\citenamefont{Martin and
  Ruostekoski}(2012{\natexlab{a}})}]{martin2012quantum}
\bibinfo{author}{\bibfnamefont{A.}~\bibnamefont{Martin}} \bibnamefont{and}
  \bibinfo{author}{\bibfnamefont{J.}~\bibnamefont{Ruostekoski}},
  \bibinfo{journal}{New J. Phys.} \textbf{\bibinfo{volume}{14}},
  \bibinfo{pages}{043040} (\bibinfo{year}{2012}{\natexlab{a}}).

\bibitem[{\citenamefont{Lewenstein and
  Malomed}(2009)}]{Lewenstein_phasekin_entangle}
\bibinfo{author}{\bibfnamefont{M.}~\bibnamefont{Lewenstein}} \bibnamefont{and}
  \bibinfo{author}{\bibfnamefont{B.~A.} \bibnamefont{Malomed}},
  \bibinfo{journal}{New J. Phys.} \textbf{\bibinfo{volume}{11}},
  \bibinfo{pages}{113014} (\bibinfo{year}{2009}).

\bibitem[{\citenamefont{Lai and Lee}(2009)}]{Lai_entanglesol_PhysRevLett}
\bibinfo{author}{\bibfnamefont{Y.}~\bibnamefont{Lai}} \bibnamefont{and}
  \bibinfo{author}{\bibfnamefont{R.-K.} \bibnamefont{Lee}},
  \bibinfo{journal}{Phys. Rev. Lett.} \textbf{\bibinfo{volume}{103}},
  \bibinfo{pages}{013902} (\bibinfo{year}{2009}).

\bibitem[{\citenamefont{Ng et~al.}(2019)\citenamefont{Ng, Opanchuk, Reid, and
  Drummond}}]{Ng_nonloc_higherorder_PhysRevLett}
\bibinfo{author}{\bibfnamefont{K.~L.} \bibnamefont{Ng}},
  \bibinfo{author}{\bibfnamefont{B.}~\bibnamefont{Opanchuk}},
  \bibinfo{author}{\bibfnamefont{M.~D.} \bibnamefont{Reid}}, \bibnamefont{and}
  \bibinfo{author}{\bibfnamefont{P.~D.} \bibnamefont{Drummond}},
  \bibinfo{journal}{Phys. Rev. Lett.} \textbf{\bibinfo{volume}{122}},
  \bibinfo{pages}{203604} (\bibinfo{year}{2019}).

\bibitem[{\citenamefont{Holdaway et~al.}(2014)\citenamefont{Holdaway, Weiss,
  and Gardiner}}]{Holdaway_entanglesol}
\bibinfo{author}{\bibfnamefont{D.~I.~H.} \bibnamefont{Holdaway}},
  \bibinfo{author}{\bibfnamefont{C.}~\bibnamefont{Weiss}}, \bibnamefont{and}
  \bibinfo{author}{\bibfnamefont{S.~A.} \bibnamefont{Gardiner}},
  \bibinfo{journal}{Phys. Rev. A} \textbf{\bibinfo{volume}{89}},
  \bibinfo{pages}{013611} (\bibinfo{year}{2014}).

\bibitem[{\citenamefont{Gertjerenken et~al.}(2013)\citenamefont{Gertjerenken,
  Billam, Blackley, Le~Sueur, Khaykovich, Cornish, and
  Weiss}}]{Gertjerenken_cat_coll_PRL}
\bibinfo{author}{\bibfnamefont{B.}~\bibnamefont{Gertjerenken}},
  \bibinfo{author}{\bibfnamefont{T.~P.} \bibnamefont{Billam}},
  \bibinfo{author}{\bibfnamefont{C.~L.} \bibnamefont{Blackley}},
  \bibinfo{author}{\bibfnamefont{C.~R.} \bibnamefont{Le~Sueur}},
  \bibinfo{author}{\bibfnamefont{L.}~\bibnamefont{Khaykovich}},
  \bibinfo{author}{\bibfnamefont{S.~L.} \bibnamefont{Cornish}},
  \bibnamefont{and} \bibinfo{author}{\bibfnamefont{C.}~\bibnamefont{Weiss}},
  \bibinfo{journal}{Phys. Rev. Lett.} \textbf{\bibinfo{volume}{111}},
  \bibinfo{pages}{100406} (\bibinfo{year}{2013}).

\bibitem[{\citenamefont{Mishmash and
  Carr}(2009)}]{Mishmash_entangleddarksol_PRL}
\bibinfo{author}{\bibfnamefont{R.~V.} \bibnamefont{Mishmash}} \bibnamefont{and}
  \bibinfo{author}{\bibfnamefont{L.~D.} \bibnamefont{Carr}},
  \bibinfo{journal}{Phys. Rev. Lett.} \textbf{\bibinfo{volume}{103}},
  \bibinfo{pages}{140403} (\bibinfo{year}{2009}).

\bibitem[{\citenamefont{Katsimiga et~al.}(2017)\citenamefont{Katsimiga,
  Koutentakis, Mistakidis, Kevrekidis, and
  Schmelcher}}]{Katsimiga_darkbright_NJP2017}
\bibinfo{author}{\bibfnamefont{G.~C.} \bibnamefont{Katsimiga}},
  \bibinfo{author}{\bibfnamefont{G.~M.} \bibnamefont{Koutentakis}},
  \bibinfo{author}{\bibfnamefont{S.~I.} \bibnamefont{Mistakidis}},
  \bibinfo{author}{\bibfnamefont{P.~G.} \bibnamefont{Kevrekidis}},
  \bibnamefont{and}
  \bibinfo{author}{\bibfnamefont{P.}~\bibnamefont{Schmelcher}},
  \bibinfo{journal}{New J. Phys.} \textbf{\bibinfo{volume}{19}},
  \bibinfo{pages}{073004} (\bibinfo{year}{2017}).

\bibitem[{\citenamefont{McGuire}(1964)}]{McGuire_exactlysolvable_JMP}
\bibinfo{author}{\bibfnamefont{J.~B.} \bibnamefont{McGuire}},
  \bibinfo{journal}{Journal of Mathematical Physics}
  \textbf{\bibinfo{volume}{5}}, \bibinfo{pages}{622} (\bibinfo{year}{1964}).

\bibitem[{\citenamefont{Lieb and Liniger}(1963)}]{LL_model_PR}
\bibinfo{author}{\bibfnamefont{E.~H.} \bibnamefont{Lieb}} \bibnamefont{and}
  \bibinfo{author}{\bibfnamefont{W.}~\bibnamefont{Liniger}},
  \bibinfo{journal}{Phys. Rev.} \textbf{\bibinfo{volume}{130}},
  \bibinfo{pages}{1605} (\bibinfo{year}{1963}).

\bibitem[{\citenamefont{Jiang et~al.}(2015)\citenamefont{Jiang, Chen, and
  Guan}}]{zhu_manybody_liebliniger_ChinPhysB}
\bibinfo{author}{\bibfnamefont{Y.-Z.} \bibnamefont{Jiang}},
  \bibinfo{author}{\bibfnamefont{Y.-Y.} \bibnamefont{Chen}}, \bibnamefont{and}
  \bibinfo{author}{\bibfnamefont{X.-W.} \bibnamefont{Guan}},
  \bibinfo{journal}{Chinese Physics B} \textbf{\bibinfo{volume}{24}},
  \bibinfo{pages}{050311} (\bibinfo{year}{2015}).

\bibitem[{\citenamefont{Bongs et~al.}(2001)\citenamefont{Bongs, Burger,
  Dettmer, Hellweg, Arlt, Ertmer, and Sengstock}}]{Bongs_waveguide_PhysRevA}
\bibinfo{author}{\bibfnamefont{K.}~\bibnamefont{Bongs}},
  \bibinfo{author}{\bibfnamefont{S.}~\bibnamefont{Burger}},
  \bibinfo{author}{\bibfnamefont{S.}~\bibnamefont{Dettmer}},
  \bibinfo{author}{\bibfnamefont{D.}~\bibnamefont{Hellweg}},
  \bibinfo{author}{\bibfnamefont{J.}~\bibnamefont{Arlt}},
  \bibinfo{author}{\bibfnamefont{W.}~\bibnamefont{Ertmer}}, \bibnamefont{and}
  \bibinfo{author}{\bibfnamefont{K.}~\bibnamefont{Sengstock}},
  \bibinfo{journal}{Phys. Rev. A} \textbf{\bibinfo{volume}{63}},
  \bibinfo{pages}{031602} (\bibinfo{year}{2001}).

\bibitem[{\citenamefont{Norrie et~al.}(2005)\citenamefont{Norrie, Ballagh, and
  Gardiner}}]{norrie:prl}
\bibinfo{author}{\bibfnamefont{A.~A.} \bibnamefont{Norrie}},
  \bibinfo{author}{\bibfnamefont{R.~J.} \bibnamefont{Ballagh}},
  \bibnamefont{and} \bibinfo{author}{\bibfnamefont{C.~W.}
  \bibnamefont{Gardiner}}, \bibinfo{journal}{Phys. Rev. Lett.}
  \textbf{\bibinfo{volume}{94}}, \bibinfo{pages}{040401}
  (\bibinfo{year}{2005}).

\bibitem[{\citenamefont{Norrie et~al.}(2006)\citenamefont{Norrie, Ballagh, and
  Gardiner}}]{norrie:long}
\bibinfo{author}{\bibfnamefont{A.~A.} \bibnamefont{Norrie}},
  \bibinfo{author}{\bibfnamefont{R.~J.} \bibnamefont{Ballagh}},
  \bibnamefont{and} \bibinfo{author}{\bibfnamefont{C.~W.}
  \bibnamefont{Gardiner}}, \bibinfo{journal}{Phys. Rev. A}
  \textbf{\bibinfo{volume}{73}}, \bibinfo{pages}{043617}
  (\bibinfo{year}{2006}).

\bibitem[{\citenamefont{Norrie}(2005)}]{norrie:thesis}
\bibinfo{author}{\bibfnamefont{A.~A.} \bibnamefont{Norrie}}, Ph.D. thesis,
  \bibinfo{school}{University of Otago} (\bibinfo{year}{2005}).

\bibitem[{\citenamefont{Gardiner and Zoller}(2004)}]{book:qn}
\bibinfo{author}{\bibfnamefont{C.~W.} \bibnamefont{Gardiner}} \bibnamefont{and}
  \bibinfo{author}{\bibfnamefont{P.}~\bibnamefont{Zoller}},
  \emph{\bibinfo{title}{Quantum Noise, 3rd ed.}}
  (\bibinfo{publisher}{Springer-Verlag, Berlin Heidelberg,},
  \bibinfo{year}{2004}).

\bibitem[{\citenamefont{Khaykovich and
  Malomed}(2006)}]{Khaykovich_Malomed_quinticsol_PRA}
\bibinfo{author}{\bibfnamefont{L.}~\bibnamefont{Khaykovich}} \bibnamefont{and}
  \bibinfo{author}{\bibfnamefont{B.~A.} \bibnamefont{Malomed}},
  \bibinfo{journal}{Phys. Rev. A} \textbf{\bibinfo{volume}{74}},
  \bibinfo{pages}{023607} (\bibinfo{year}{2006}).

\bibitem[{\citenamefont{Jisha et~al.}(2015)\citenamefont{Jisha, Mithun,
  Rodrigues, and Porsezian}}]{Jisha:15_solitons_josab}
\bibinfo{author}{\bibfnamefont{C.~P.} \bibnamefont{Jisha}},
  \bibinfo{author}{\bibfnamefont{T.}~\bibnamefont{Mithun}},
  \bibinfo{author}{\bibfnamefont{A.}~\bibnamefont{Rodrigues}},
  \bibnamefont{and}
  \bibinfo{author}{\bibfnamefont{K.}~\bibnamefont{Porsezian}},
  \bibinfo{journal}{J. Opt. Soc. Am. B} \textbf{\bibinfo{volume}{32}},
  \bibinfo{pages}{1106} (\bibinfo{year}{2015}).

\bibitem[{\citenamefont{Kivshar and Agrawal}(2003)}]{kivshar2003optical}
\bibinfo{author}{\bibfnamefont{Y.~S.} \bibnamefont{Kivshar}} \bibnamefont{and}
  \bibinfo{author}{\bibfnamefont{G.}~\bibnamefont{Agrawal}},
  \emph{\bibinfo{title}{Optical solitons: from fibers to photonic crystals}}
  (\bibinfo{publisher}{Academic press}, \bibinfo{year}{2003}).

\bibitem[{\citenamefont{Zegadlo et~al.}(2014)\citenamefont{Zegadlo, Wasak,
  Malomed, Karpierz, and Trippenbach}}]{malomed_stable}
\bibinfo{author}{\bibfnamefont{K.~B.} \bibnamefont{Zegadlo}},
  \bibinfo{author}{\bibfnamefont{T.}~\bibnamefont{Wasak}},
  \bibinfo{author}{\bibfnamefont{B.~A.} \bibnamefont{Malomed}},
  \bibinfo{author}{\bibfnamefont{M.~A.} \bibnamefont{Karpierz}},
  \bibnamefont{and}
  \bibinfo{author}{\bibfnamefont{M.}~\bibnamefont{Trippenbach}},
  \bibinfo{journal}{Chaos: An Interdisciplinary Journal of Nonlinear Science}
  \textbf{\bibinfo{volume}{24}}, \bibinfo{pages}{043136}
  (\bibinfo{year}{2014}).

\bibitem[{\citenamefont{Nath et~al.}(2017)\citenamefont{Nath, Roy, and
  Roychoudhury}}]{nath_quintic}
\bibinfo{author}{\bibfnamefont{D.}~\bibnamefont{Nath}},
  \bibinfo{author}{\bibfnamefont{B.}~\bibnamefont{Roy}}, \bibnamefont{and}
  \bibinfo{author}{\bibfnamefont{R.}~\bibnamefont{Roychoudhury}},
  \bibinfo{journal}{Optics Communications} \textbf{\bibinfo{volume}{393}},
  \bibinfo{pages}{224} (\bibinfo{year}{2017}).

\bibitem[{\citenamefont{Lee et~al.}(2004)\citenamefont{Lee, Lai, and
  Malomed}}]{lee2004quantum}
\bibinfo{author}{\bibfnamefont{R.-K.} \bibnamefont{Lee}},
  \bibinfo{author}{\bibfnamefont{Y.}~\bibnamefont{Lai}}, \bibnamefont{and}
  \bibinfo{author}{\bibfnamefont{B.~A.} \bibnamefont{Malomed}},
  \bibinfo{journal}{Journal of Optics B: Quantum and Semiclassical Optics}
  \textbf{\bibinfo{volume}{6}}, \bibinfo{pages}{367} (\bibinfo{year}{2004}).

\bibitem[{\citenamefont{Baizakov et~al.}(2019)\citenamefont{Baizakov, Bouketir,
  Al-Marzoug, and Bahlouli}}]{baizakov2019effect}
\bibinfo{author}{\bibfnamefont{B.}~\bibnamefont{Baizakov}},
  \bibinfo{author}{\bibfnamefont{A.}~\bibnamefont{Bouketir}},
  \bibinfo{author}{\bibfnamefont{S.}~\bibnamefont{Al-Marzoug}},
  \bibnamefont{and} \bibinfo{author}{\bibfnamefont{H.}~\bibnamefont{Bahlouli}},
  \bibinfo{journal}{Optik} \textbf{\bibinfo{volume}{180}}, \bibinfo{pages}{792}
  (\bibinfo{year}{2019}).

\bibitem[{\citenamefont{Gordon}(1983)}]{gordon_forces}
\bibinfo{author}{\bibfnamefont{J.~P.} \bibnamefont{Gordon}},
  \bibinfo{journal}{Opt. Lett.} \textbf{\bibinfo{volume}{8}},
  \bibinfo{pages}{596} (\bibinfo{year}{1983}).

\bibitem[{\citenamefont{Al~Khawaja et~al.}(2002)\citenamefont{Al~Khawaja,
  Stoof, Hulet, Strecker, and Partridge}}]{stoof_solitons}
\bibinfo{author}{\bibfnamefont{U.}~\bibnamefont{Al~Khawaja}},
  \bibinfo{author}{\bibfnamefont{H.~T.~C.} \bibnamefont{Stoof}},
  \bibinfo{author}{\bibfnamefont{R.~G.} \bibnamefont{Hulet}},
  \bibinfo{author}{\bibfnamefont{K.~E.} \bibnamefont{Strecker}},
  \bibnamefont{and} \bibinfo{author}{\bibfnamefont{G.~B.}
  \bibnamefont{Partridge}}, \bibinfo{journal}{Phys. Rev. Lett.}
  \textbf{\bibinfo{volume}{89}}, \bibinfo{pages}{200404}
  (\bibinfo{year}{2002}).

\bibitem[{\citenamefont{Cowan et~al.}(1986)\citenamefont{Cowan, Enns,
  Rangnekar, and Sanghera}}]{Cowan_QuasiSoliton_CJP_1986}
\bibinfo{author}{\bibfnamefont{S.}~\bibnamefont{Cowan}},
  \bibinfo{author}{\bibfnamefont{R.~H.} \bibnamefont{Enns}},
  \bibinfo{author}{\bibfnamefont{S.~S.} \bibnamefont{Rangnekar}},
  \bibnamefont{and} \bibinfo{author}{\bibfnamefont{S.~S.}
  \bibnamefont{Sanghera}}, \bibinfo{journal}{Canadian Journal of Physics}
  \textbf{\bibinfo{volume}{64}}, \bibinfo{pages}{311} (\bibinfo{year}{1986}).

\bibitem[{\citenamefont{{Sergio Bezerra
  Sombra}}(1992)}]{Sergio_Bistable_OC1992}
\bibinfo{author}{\bibfnamefont{A.}~\bibnamefont{{Sergio Bezerra Sombra}}},
  \bibinfo{journal}{Optics Communications} \textbf{\bibinfo{volume}{94}},
  \bibinfo{pages}{92} (\bibinfo{year}{1992}).

\bibitem[{\citenamefont{Soneson and
  Peleg}(2004)}]{Soneson_quinticOpticalfibres_PhysicaD}
\bibinfo{author}{\bibfnamefont{J.}~\bibnamefont{Soneson}} \bibnamefont{and}
  \bibinfo{author}{\bibfnamefont{A.}~\bibnamefont{Peleg}},
  \bibinfo{journal}{Physica D} \textbf{\bibinfo{volume}{195}},
  \bibinfo{pages}{123} (\bibinfo{year}{2004}).

\bibitem[{\citenamefont{Konar et~al.}(2006)\citenamefont{Konar, Mishra, and
  Jana}}]{Konar_quinticPropagation_CSF2006}
\bibinfo{author}{\bibfnamefont{S.}~\bibnamefont{Konar}},
  \bibinfo{author}{\bibfnamefont{M.}~\bibnamefont{Mishra}}, \bibnamefont{and}
  \bibinfo{author}{\bibfnamefont{S.}~\bibnamefont{Jana}},
  \bibinfo{journal}{Chaos, Solitons \& Fractals} \textbf{\bibinfo{volume}{29}},
  \bibinfo{pages}{823} (\bibinfo{year}{2006}).

\bibitem[{\citenamefont{Xie et~al.}(2016)\citenamefont{Xie, Tian, Sun, Liu, and
  Jiang}}]{Xie2016_BrightSolitons_Optics_OQE}
\bibinfo{author}{\bibfnamefont{X.-Y.} \bibnamefont{Xie}},
  \bibinfo{author}{\bibfnamefont{B.}~\bibnamefont{Tian}},
  \bibinfo{author}{\bibfnamefont{Y.}~\bibnamefont{Sun}},
  \bibinfo{author}{\bibfnamefont{L.}~\bibnamefont{Liu}}, \bibnamefont{and}
  \bibinfo{author}{\bibfnamefont{Y.}~\bibnamefont{Jiang}},
  \bibinfo{journal}{Optical and Quantum Electronics}
  \textbf{\bibinfo{volume}{48}}, \bibinfo{pages}{1} (\bibinfo{year}{2016}).

\bibitem[{\citenamefont{Albuch and
  Malomed}(2007)}]{Albuch_SymAsymOptics_MATCOM_2007}
\bibinfo{author}{\bibfnamefont{L.}~\bibnamefont{Albuch}} \bibnamefont{and}
  \bibinfo{author}{\bibfnamefont{B.~A.} \bibnamefont{Malomed}},
  \bibinfo{journal}{Mathematics and Computers in Simulation}
  \textbf{\bibinfo{volume}{74}}, \bibinfo{pages}{312} (\bibinfo{year}{2007}).

\bibitem[{\citenamefont{Lewenstein and You}(1996)}]{lewenstein_phasediff}
\bibinfo{author}{\bibfnamefont{M.}~\bibnamefont{Lewenstein}} \bibnamefont{and}
  \bibinfo{author}{\bibfnamefont{L.}~\bibnamefont{You}},
  \bibinfo{journal}{Phys. Rev. Lett.} \textbf{\bibinfo{volume}{77}},
  \bibinfo{pages}{3489} (\bibinfo{year}{1996}).

\bibitem[{\citenamefont{Streltsov et~al.}(2011)\citenamefont{Streltsov, Alon,
  and Cederbaum}}]{streltsov_frag}
\bibinfo{author}{\bibfnamefont{A.~I.} \bibnamefont{Streltsov}},
  \bibinfo{author}{\bibfnamefont{O.~E.} \bibnamefont{Alon}}, \bibnamefont{and}
  \bibinfo{author}{\bibfnamefont{L.~S.} \bibnamefont{Cederbaum}},
  \bibinfo{journal}{Phys. Rev. Lett.} \textbf{\bibinfo{volume}{106}},
  \bibinfo{pages}{240401} (\bibinfo{year}{2011}).

\bibitem[{\citenamefont{Sreedharan et~al.}(2020)\citenamefont{Sreedharan,
  Choudhury, Mukherjee, Streltsov, and
  W{\"u}ster}}]{Aparna_collisions_beyondMT_PRA}
\bibinfo{author}{\bibfnamefont{A.}~\bibnamefont{Sreedharan}},
  \bibinfo{author}{\bibfnamefont{S.}~\bibnamefont{Choudhury}},
  \bibinfo{author}{\bibfnamefont{R.}~\bibnamefont{Mukherjee}},
  \bibinfo{author}{\bibfnamefont{A.}~\bibnamefont{Streltsov}},
  \bibnamefont{and}
  \bibinfo{author}{\bibfnamefont{S.}~\bibnamefont{W{\"u}ster}},
  \bibinfo{journal}{Phys. Rev. A} \textbf{\bibinfo{volume}{101}},
  \bibinfo{pages}{043604} (\bibinfo{year}{2020}).

\bibitem[{\citenamefont{Dennis et~al.}(2013)\citenamefont{Dennis, Hope, and
  Johnsson}}]{xmds:paper}
\bibinfo{author}{\bibfnamefont{G.~R.} \bibnamefont{Dennis}},
  \bibinfo{author}{\bibfnamefont{J.~J.} \bibnamefont{Hope}}, \bibnamefont{and}
  \bibinfo{author}{\bibfnamefont{M.~T.} \bibnamefont{Johnsson}},
  \bibinfo{journal}{Comput. Phys. Comm.} \textbf{\bibinfo{volume}{184}},
  \bibinfo{pages}{201} (\bibinfo{year}{2013}).

\bibitem[{\citenamefont{Dennis et~al.}(2012)\citenamefont{Dennis, Hope, and
  Johnsson}}]{xmds:docu}
\bibinfo{author}{\bibfnamefont{G.~R.} \bibnamefont{Dennis}},
  \bibinfo{author}{\bibfnamefont{J.~J.} \bibnamefont{Hope}}, \bibnamefont{and}
  \bibinfo{author}{\bibfnamefont{M.~T.} \bibnamefont{Johnsson}}
  (\bibinfo{year}{2012}), \bibinfo{note}{http://www.xmds.org/}.

\bibitem[{\citenamefont{Papacharalampous
  et~al.}(2003)\citenamefont{Papacharalampous, Kevrekidis, Malomed, and
  Frantzeskakis}}]{Papacharalampous_solcoll_DNLSE}
\bibinfo{author}{\bibfnamefont{I.~E.} \bibnamefont{Papacharalampous}},
  \bibinfo{author}{\bibfnamefont{P.~G.} \bibnamefont{Kevrekidis}},
  \bibinfo{author}{\bibfnamefont{B.~A.} \bibnamefont{Malomed}},
  \bibnamefont{and} \bibinfo{author}{\bibfnamefont{D.~J.}
  \bibnamefont{Frantzeskakis}}, \bibinfo{journal}{Phys. Rev. E}
  \textbf{\bibinfo{volume}{68}}, \bibinfo{pages}{046604}
  (\bibinfo{year}{2003}).

\bibitem[{\citenamefont{Weiss et~al.}(2015)\citenamefont{Weiss, Gardiner, and
  Breuer}}]{weiss_CMdiffusion}
\bibinfo{author}{\bibfnamefont{C.}~\bibnamefont{Weiss}},
  \bibinfo{author}{\bibfnamefont{S.~A.} \bibnamefont{Gardiner}},
  \bibnamefont{and} \bibinfo{author}{\bibfnamefont{H.-P.}
  \bibnamefont{Breuer}}, \bibinfo{journal}{Phys. Rev. A}
  \textbf{\bibinfo{volume}{91}}, \bibinfo{pages}{063616}
  (\bibinfo{year}{2015}).

\bibitem[{\citenamefont{Cosme et~al.}(2016)\citenamefont{Cosme, Weiss, and
  Brand}}]{Cosme_com_motion}
\bibinfo{author}{\bibfnamefont{J.~G.} \bibnamefont{Cosme}},
  \bibinfo{author}{\bibfnamefont{C.}~\bibnamefont{Weiss}}, \bibnamefont{and}
  \bibinfo{author}{\bibfnamefont{J.}~\bibnamefont{Brand}},
  \bibinfo{journal}{Phys. Rev. A} \textbf{\bibinfo{volume}{94}},
  \bibinfo{pages}{043603} (\bibinfo{year}{2016}).

\bibitem[{\citenamefont{Sreedharan et~al.}(2022)\citenamefont{Sreedharan,
  Sridevi, Choudhury, Mukherjee, Streltsov, and
  W{\"u}ster}}]{Sreedharan_backforth}
\bibinfo{author}{\bibfnamefont{A.}~\bibnamefont{Sreedharan}},
  \bibinfo{author}{\bibfnamefont{K.}~\bibnamefont{Sridevi}},
  \bibinfo{author}{\bibfnamefont{S.}~\bibnamefont{Choudhury}},
  \bibinfo{author}{\bibfnamefont{R.}~\bibnamefont{Mukherjee}},
  \bibinfo{author}{\bibfnamefont{A.}~\bibnamefont{Streltsov}},
  \bibnamefont{and}
  \bibinfo{author}{\bibfnamefont{S.}~\bibnamefont{W{\"u}ster}}
  (\bibinfo{year}{2022}), \bibinfo{note}{arXiv:1904.06552}.

\bibitem[{\citenamefont{Zakharov and Shabat}(1972)}]{Zakharov_solution_NLSE}
\bibinfo{author}{\bibfnamefont{V.}~\bibnamefont{Zakharov}} \bibnamefont{and}
  \bibinfo{author}{\bibfnamefont{A.}~\bibnamefont{Shabat}},
  \bibinfo{journal}{Sov. Phys. JETP-Ussr} \textbf{\bibinfo{volume}{34}},
  \bibinfo{pages}{62} (\bibinfo{year}{1972}).

\bibitem[{\citenamefont{Kinoshita et~al.}(2006)\citenamefont{Kinoshita, Wenger,
  and Weiss}}]{cradle}
\bibinfo{author}{\bibfnamefont{T.}~\bibnamefont{Kinoshita}},
  \bibinfo{author}{\bibfnamefont{T.}~\bibnamefont{Wenger}}, \bibnamefont{and}
  \bibinfo{author}{\bibfnamefont{D.~S.} \bibnamefont{Weiss}},
  \bibinfo{journal}{Nature} \textbf{\bibinfo{volume}{440}},
  \bibinfo{pages}{900} (\bibinfo{year}{2006}).

\bibitem[{\citenamefont{Lai and Haus}(1989)}]{lai_quantsol_II}
\bibinfo{author}{\bibfnamefont{Y.}~\bibnamefont{Lai}} \bibnamefont{and}
  \bibinfo{author}{\bibfnamefont{H.~A.} \bibnamefont{Haus}},
  \bibinfo{journal}{Phys. Rev. A} \textbf{\bibinfo{volume}{40}},
  \bibinfo{pages}{854} (\bibinfo{year}{1989}).

\bibitem[{\citenamefont{Karthik et~al.}(2007)\citenamefont{Karthik, Sharma, and
  Lakshminarayan}}]{Karthik_entangle_chaos_PRA}
\bibinfo{author}{\bibfnamefont{J.}~\bibnamefont{Karthik}},
  \bibinfo{author}{\bibfnamefont{A.}~\bibnamefont{Sharma}}, \bibnamefont{and}
  \bibinfo{author}{\bibfnamefont{A.}~\bibnamefont{Lakshminarayan}},
  \bibinfo{journal}{Phys. Rev. A} \textbf{\bibinfo{volume}{75}},
  \bibinfo{pages}{022304} (\bibinfo{year}{2007}).

\bibitem[{\citenamefont{Deutsch et~al.}(2013)\citenamefont{Deutsch, Li, and
  Sharma}}]{Deutsch_microsc_entropy_PhysRevE}
\bibinfo{author}{\bibfnamefont{J.~M.} \bibnamefont{Deutsch}},
  \bibinfo{author}{\bibfnamefont{H.}~\bibnamefont{Li}}, \bibnamefont{and}
  \bibinfo{author}{\bibfnamefont{A.}~\bibnamefont{Sharma}},
  \bibinfo{journal}{Phys. Rev. E} \textbf{\bibinfo{volume}{87}},
  \bibinfo{pages}{042135} (\bibinfo{year}{2013}).

\bibitem[{sup()}]{sup:info}
\bibinfo{note}{See Supplemental Material at [URL will be inserted by publisher]
  for more single TWA trajectories, our algorithm for quantum noise removal
  from soliton velocities and entanglement criteria for pure states.}

\bibitem[{foo()}]{footnote:TMMcoeffs}
\bibinfo{note}{Specifically $\chi= \sub{g}{1D} \int dx
  |L(x)|^4=-6.56036\times10^{-37}$, $\eta= -\sub{q}{2} \int dx
  |L(x)|^6=-5.32715\times10^{-42}<0$.}

\bibitem[{\citenamefont{Martin and
  Ruostekoski}(2012{\natexlab{b}})}]{Martin_solinterf_NJP_2012}
\bibinfo{author}{\bibfnamefont{A.~D.} \bibnamefont{Martin}} \bibnamefont{and}
  \bibinfo{author}{\bibfnamefont{J.}~\bibnamefont{Ruostekoski}},
  \bibinfo{journal}{New J. Phys.} \textbf{\bibinfo{volume}{14}},
  \bibinfo{pages}{043040} (\bibinfo{year}{2012}{\natexlab{b}}).

\bibitem[{\citenamefont{Drummond and Gardiner}(1980)}]{drummond1980generalised}
\bibinfo{author}{\bibfnamefont{P.}~\bibnamefont{Drummond}} \bibnamefont{and}
  \bibinfo{author}{\bibfnamefont{C.}~\bibnamefont{Gardiner}},
  \bibinfo{journal}{Journal of Physics A: Mathematical and General}
  \textbf{\bibinfo{volume}{13}}, \bibinfo{pages}{2353} (\bibinfo{year}{1980}).

\bibitem[{\citenamefont{Deuar and
  Drummond}(2006{\natexlab{a}})}]{deuar2006first}
\bibinfo{author}{\bibfnamefont{P.}~\bibnamefont{Deuar}} \bibnamefont{and}
  \bibinfo{author}{\bibfnamefont{P.}~\bibnamefont{Drummond}},
  \bibinfo{journal}{Journal of Physics A: Mathematical and General}
  \textbf{\bibinfo{volume}{39}}, \bibinfo{pages}{1163}
  (\bibinfo{year}{2006}{\natexlab{a}}).

\bibitem[{\citenamefont{Deuar and
  Drummond}(2006{\natexlab{b}})}]{deuar2006second}
\bibinfo{author}{\bibfnamefont{P.}~\bibnamefont{Deuar}} \bibnamefont{and}
  \bibinfo{author}{\bibfnamefont{P.}~\bibnamefont{Drummond}},
  \bibinfo{journal}{Journal of Physics A: Mathematical and General}
  \textbf{\bibinfo{volume}{39}}, \bibinfo{pages}{2723}
  (\bibinfo{year}{2006}{\natexlab{b}}).

\bibitem[{\citenamefont{Zeitler et~al.}(2022)\citenamefont{Zeitler, Chapman,
  Chitambar, and Kwiat}}]{Zeitler_hyperent_arxiv}
\bibinfo{author}{\bibfnamefont{C.~K.} \bibnamefont{Zeitler}},
  \bibinfo{author}{\bibfnamefont{J.~C.} \bibnamefont{Chapman}},
  \bibinfo{author}{\bibfnamefont{E.}~\bibnamefont{Chitambar}},
  \bibnamefont{and} \bibinfo{author}{\bibfnamefont{P.~G.} \bibnamefont{Kwiat}}
  (\bibinfo{year}{2022}), \bibinfo{note}{arXiv:2207.09990}.

\bibitem[{\citenamefont{Benatti et~al.}(2020)\citenamefont{Benatti, Floreanini,
  Franchini, and Marzolino}}]{Benatti_indist_entangle_review}
\bibinfo{author}{\bibfnamefont{F.}~\bibnamefont{Benatti}},
  \bibinfo{author}{\bibfnamefont{R.}~\bibnamefont{Floreanini}},
  \bibinfo{author}{\bibfnamefont{F.}~\bibnamefont{Franchini}},
  \bibnamefont{and}
  \bibinfo{author}{\bibfnamefont{U.}~\bibnamefont{Marzolino}},
  \bibinfo{journal}{Physics Reports} \textbf{\bibinfo{volume}{878}},
  \bibinfo{pages}{1} (\bibinfo{year}{2020}).

\bibitem[{\citenamefont{Simon}(2000)}]{Simon_entanglecrit_PhysRevLett}
\bibinfo{author}{\bibfnamefont{R.}~\bibnamefont{Simon}},
  \bibinfo{journal}{Phys. Rev. Lett.} \textbf{\bibinfo{volume}{84}},
  \bibinfo{pages}{2726} (\bibinfo{year}{2000}).

\bibitem[{\citenamefont{Tan}(1999)}]{quantele_PRA}
\bibinfo{author}{\bibfnamefont{S.~M.} \bibnamefont{Tan}},
  \bibinfo{journal}{Phys. Rev. A} \textbf{\bibinfo{volume}{60}},
  \bibinfo{pages}{2752} (\bibinfo{year}{1999}).

\bibitem[{\citenamefont{Cirac et~al.}(1998)\citenamefont{Cirac, Lewenstein,
  M\o{}lmer, and Zoller}}]{Cirac_superposBEC_PhysRevA}
\bibinfo{author}{\bibfnamefont{J.~I.} \bibnamefont{Cirac}},
  \bibinfo{author}{\bibfnamefont{M.}~\bibnamefont{Lewenstein}},
  \bibinfo{author}{\bibfnamefont{K.}~\bibnamefont{M\o{}lmer}},
  \bibnamefont{and} \bibinfo{author}{\bibfnamefont{P.}~\bibnamefont{Zoller}},
  \bibinfo{journal}{Phys. Rev. A} \textbf{\bibinfo{volume}{57}},
  \bibinfo{pages}{1208} (\bibinfo{year}{1998}).

\bibitem[{\citenamefont{Weiss and Castin}(2009)}]{weiss:solitoncat}
\bibinfo{author}{\bibfnamefont{C.}~\bibnamefont{Weiss}} \bibnamefont{and}
  \bibinfo{author}{\bibfnamefont{Y.}~\bibnamefont{Castin}},
  \bibinfo{journal}{Phys. Rev. Lett.} \textbf{\bibinfo{volume}{102}},
  \bibinfo{pages}{010403} (\bibinfo{year}{2009}).

\bibitem[{\citenamefont{M{\"o}bius et~al.}(2013)\citenamefont{M{\"o}bius,
  Genkin, Eisfeld, W{\"u}ster, and {J.-M. Rost}}}]{moebius:cat}
\bibinfo{author}{\bibfnamefont{S.}~\bibnamefont{M{\"o}bius}},
  \bibinfo{author}{\bibfnamefont{M.}~\bibnamefont{Genkin}},
  \bibinfo{author}{\bibfnamefont{A.}~\bibnamefont{Eisfeld}},
  \bibinfo{author}{\bibfnamefont{S.}~\bibnamefont{W{\"u}ster}},
  \bibnamefont{and} \bibinfo{author}{\bibnamefont{{J.-M. Rost}}},
  \bibinfo{journal}{Phys. Rev. A} \textbf{\bibinfo{volume}{87}},
  \bibinfo{pages}{051602(R)} (\bibinfo{year}{2013}).

\bibitem[{\citenamefont{D\c{a}browska-W{\"u}ster
  et~al.}(2009)\citenamefont{D\c{a}browska-W{\"u}ster, W{\"u}ster, and
  Davis}}]{wuester:collsoll}
\bibinfo{author}{\bibfnamefont{B.~J.} \bibnamefont{D\c{a}browska-W{\"u}ster}},
  \bibinfo{author}{\bibfnamefont{S.}~\bibnamefont{W{\"u}ster}},
  \bibnamefont{and} \bibinfo{author}{\bibfnamefont{M.~J.} \bibnamefont{Davis}},
  \bibinfo{journal}{New J. Phys.} \textbf{\bibinfo{volume}{11}},
  \bibinfo{pages}{053017} (\bibinfo{year}{2009}).

\bibitem[{\citenamefont{Carr and Brand}(2004)}]{brand_solitons}
\bibinfo{author}{\bibfnamefont{L.~D.} \bibnamefont{Carr}} \bibnamefont{and}
  \bibinfo{author}{\bibfnamefont{J.}~\bibnamefont{Brand}},
  \bibinfo{journal}{Phys. Rev. Lett.} \textbf{\bibinfo{volume}{92}},
  \bibinfo{pages}{040401} (\bibinfo{year}{2004}).

\end{thebibliography}




\end{document}